\documentclass[twocolumn]{aa}
\usepackage{graphicx}
\usepackage{psfig}

\usepackage{txfonts}
\newcommand{\clvier}{Cl\,0413--65}
\newcommand{\mseins}{MS\,1008--12}
\newcommand{\cldrei}{Cl\,0303$+$17}
\newcommand{\science}{Paper\,I}

\begin{document}
   \title{Internal kinematics of spiral galaxies in distant clusters. Part II.
   \thanks{Based on observations collected at the European Southern 
   Observatory (ESO), Cerro Paranal, Chile (ESO Nos. 64.O--0158, 64.O--0152 \& 
   66.A--0547)}}

   \subtitle{Observations and data analysis.}

   \author{K. J\"ager\inst{1}, B. L. Ziegler\inst{1}, A. B\"ohm\inst{1}
          J. Heidt\inst{2}, C.~M\"ollenhoff\inst{2}, U.~Hopp\inst{3}, 
	  R.H.~Mendez\inst{4}, and S.~Wagner\inst{2}.
          }

   \offprints{K. J\"ager}

   \institute{Universit\"atssternwarte G\"ottingen, Geismarlandstr.~11, 37083
              G\"ottingen, Germany\\
              \email{jaeger@uni-sw.gwdg.de}
         \and
             Landessternwarte Heidelberg, K\"onigstuhl, D--69117 Heidelberg, 
	     Germany
         \and 
	     Universit\"atssternwarte M\"unchen, Scheinerstr.~1, 81679 
	     M\"unchen, Germany
	 \and 
	     Institute for Astronomy,2680 Woodlawn Drive, Honolulu, Hawaii 96822, USA  
	     }

   \date{Received; accepted}

   \abstract{We have conducted an observing campaign with the FORS instruments 
   at the ESO--Very Large Telescope 
   to explore the kinematical properties of spiral galaxies in distant 
   galaxy clusters. Our
   main goal is to analyse transformation-- and interaction processes of disk 
   galaxies within the special environment of clusters as compared to the 
   hierarchical evolution of galaxies in the field. 
   Spatially resolved multi object spectra
   have been obtained for seven galaxy clusters at 0.3$<$z$<$0.6 
   to measure rotation velocities of cluster members. 
   For three of the clusters, \cldrei, \clvier, and \mseins, 
   for which we presented results including a Tully--Fisher--diagram
   in Ziegler et al. (\cite{Ziegler}),    
   we describe here in detail the sample selection, observations, data
   reduction, and data analysis. 
   Each of them was observed with two setups of the standard MOS--unit of 
   FORS. With typical exposure times of $>$2 hours we reach an 
   S/N$>$5 in the emission lines appropriate for the deduction of the galaxies'
   internal rotation velocities 
   from [OII], $H_{\beta}$, or [OIII] emission line profiles. 
   Preselection of targets was done on the basis of available 
   redshifts as well as from photometric and morphological information gathered 
   from own observations, archive data, and from the literature. Emphasis was
   laid on the definition of suitable setups to avoid the typical 
   restrictions of the standard MOS unit for this kind of observations.
   In total we assembled spectra of 116 objects of which 50 turned out to be 
   cluster members. 
   Position velocity diagrams, finding charts for each setup as well as tables 
   with photometric, spectral, and structural parameters of individual galaxies 
   are presented.
  
   }
   \titlerunning{Internal kinematics of spiral galaxies in distant clusters. 
   Part II. Observations and data analysis.}
   \authorrunning{K. J\"ager et al.}
   \maketitle
%
\section{Introduction}
\begin{table*}
\centering
\caption{Observation log of our MOS-- and imaging data. 
}
\begin{tabular}{lcccccclc}
\hline
Cluster & RA (2000) & DEC (2000) & z & Setup & Config. & Date & Exposures & FWHM [arcsec] \\
\hline

Cl\,0303$+$1706  & 03 06 18.7 & +17 19 22 & 0.42 &A&  MOS/FORS2 &10/2000 & $3\times2400$s      &0.43\\
                 &            &	          &      &B&  MOS/FORS2 &10/2000 & $3\times2400$s      &0.90\\
		 &            &           &      & &  IMA/FORS1 &10/1999 & $3\times300$s in R  &1.00\\
\hline
Cl\,0413--6559   & 04 12 54.7 & -65 50 58 & 0.51 &A&  MOS/FORS1 &11/1999 & $4\times1800$s      &1.33\\
                 &            &	          &      &B&  MOS/FORS1 &11/1999 & $3\times2250$s      &0.72\\
		 &            &	          &      & &            &        & $2\times1800$s      &0.61\\
		 &            &	          &      & &  IMA/FORS1 &12/1998 & $600$s in I         &0.77\\
		 &            &	          &      & &  IMA/FORS1 &12/1998 & $600$s in R         &0.84\\
\hline
MS\,1008.1--1224 & 10 10 34.1 &-12 39 48  & 0.30 &A&  MOS/FORS1 &03/2000 & $3\times2700$s      &0.70\\
                 &            &	          &      &B&  MOS/FORS1 &03/2000 & $3\times2700$s      &0.70\\
                 &            &	          &      & &  IMA/FORS2 &02/2000 & $300$s in I         &0.59\\
\hline
\end{tabular}
\end{table*}
With the advent of 10m--class telescopes we are now able to analyse the 
internal kinematics of disk galaxies as a function of density environment 
over a wide range of cosmic epochs.
Spatially resolved galaxy spectra up to redshifts $z\approx 1$ are needed to 
compare, for example, the Tully--Fisher--Relation (TFR, Tully \& Fisher 
\cite{TF77}) of local galaxies in clusters and the field with the TFR of their 
more distant and therefore much younger counterparts. 

Following models of hierarchically growing structure, clusters of 
galaxies are still in the process of forming at $z=1$ in the 
concordance cosmology. High resolution Hubble Space Telecope 
(HST) and ground--based images indicate that the morphological structure of 
cluster galaxies is affected by different phenomena related to their dense
environment. This implies a higher galaxy infall rate, 
star formation rate and frequency of interactions of the cluster members at 
intermediate redshifts (e.g.~Kodama and Bower \cite{KB01}).
However, any specific evolution of cluster galaxies is superposed on the
hierarchical evolution of galaxies which is characterized by the growth of 
objects, by mergers, and declining star formation rates.
Furthermore, cluster members should interact with  
the intracluster medium which fills the 
gravitational potential of clusters. Finally, it is possible that the 
the dark matter halo of cluster galaxies (and therefore their 
total mass) is undergoing distinct changes by particular 
interaction processes.\\
One important scenario presently discussed came from the finding that
local clusters are dominated by elliptical and lenticular galaxies 
while the distant ones show a high fraction of spiral and irregular 
galaxies (Dressler et al. \cite{Dressler}). Spirals from the field might fall 
into clusters and experience a morphological transformation into S0 galaxies. 
Some conceivable interactions, like merging, 
should cause substantial distortions of the internal kinematics of 
disk galaxies with velocity profiles no longer following the universal form
(Persic et al. \cite{PS96}). 
Galaxies with ''regular'' rotation curves can be used to investigate their 
luminosity evolution via the TFR which connects their internal kinematics 
to their stellar population.
First studies of distant clusters have been done by 
Milvang--Jensen et al. 
(\cite{MJ}) (7 TF spiral galaxies within MS 1054.4$-$0321 at $z=0.83$) and   
Metevier et al. (\cite{Metevier}) (10 TF spirals within CL0024+1654 at 
$z=0.40$) using the VLT and the Keck telescope, respectively.
In Ziegler et al. (\cite{Ziegler}, hereafter Paper I) we recently presented 
first results of our FORS--VLT MOS observations of three clusters  
at intermediate redshift. Both position velocity diagrams of
late--type galaxies as 
well as a TF--diagram for 13 cluster members were shown supplemented 
by a comparison of our results with those from the publications mentioned above.
In this paper we focus on a more detailed description of 
our observations and measurements and present the data in tabular form.

Our overall project comprises  
observations of galaxies in the fields of seven rich clusters 
within the redshift range $0.3<z<0.6$. 
From spatially resolved spectra we investigate internal kinematics.
Photometric measurements complete the data set.
All observations have been carried out with the two 
FORS instruments ({\bf{FO}}cal {\bf{R}}educer and {\bf{S}}pectrograph) 
mounted at the Very Large Telescope (VLT) of the European Southern 
Observatory (ESO) (see  
Appenzeller et al. (\cite{Appenzeller}) for a comprehensive list of technical 
papers on FORS). 
While FORS2 now supports exchangable masks ({\bf{M}}ask {\bf{E}}xchange 
{\bf{U}}nit -- MXU)\footnote{with individual laser--cut slitlets}, 
all observations of \clvier\ , \mseins\  and 
\cldrei\ were still restricted to the 
standard Multi Object Spectroscopy Unit (MOS) with 19 moveable 
slitlets. Only these observations (Table 1) are discussed 
within this paper. The other four clusters will be 
discussed in a future publication since different reduction techniques are
necessary.

The paper is organized as follows:
Section two describes our imaging observations and the determination of 
photometric values and absolute magnitudes which are important for the 
construction of a TF--diagram.
Section three gives an overview about the target selection for our 
MOS observations and how we deal with some 
restrictions of the standard MOS unit of FORS.
Section four describes the observations and reduction of the MOS data, 
while section five explains our measurements of internal galaxy 
kinematics. In section six finding charts, position velocity diagrams 
and data tables are presented. 
\section{Imaging and photometry}
Direct imaging of the cluster fields was needed to prepare 
the MOS setups and to obtain photometric and structural 
information of the galaxies for the TF--diagram.

For \cldrei\ three R band FORS1 images with 300s exposure time each were 
taken in October 1999. Photometric information for \clvier\ was gathered from 
two FORS1 images in R and I with 600s exposure time, respectively,
taken in December 1998. 
For \mseins\ we could use the FORS Cluster Deep Field images 
(see Lombardi et al. \cite{Lom}). Additionally, we took an acquisition image 
of 300s in I in February 2000 with FORS2.

Except for the Cluster Deep Field images which were available from the 
ESO archive\footnote{http://www.eso.org/science/ut1sv/Clus\_index.html} 
already reduced and calibrated,
all own FORS images were reduced with standard procedures of  
ESO--MIDAS\footnote{MIDAS: {\bf{M}}unich {\bf{I}}mage {\bf{D}}ata 
{\bf{A}}nalysis {\bf{S}}ystem}.
Each science and flatfield image was firstly 
bias substracted by using a master bias which was created 
from a stack of single bias frames and which was scaled to the overscan level 
of the individual image. Skyflats were taken to 
correct for the pixel to pixel variations whereas the 
large scale gradients were removed by a normalized superflat 
created from a medianed and smoothed stack of further 
science frames taken during the same nights. 
Finally,  tracks of cosmic rays have been removed
with a standard MIDAS procedure (see section 4 for some details).
Photometric zeropoints were calculated from observations of Landolt 
standard fields (Landolt \cite{Landolt}) during the same nights. 

The photometry of the galaxies was performed with the Source Extractor package 
(SExtractor) by Bertin \& Arnouts (\cite{Bertin}).
As total apparent magnitudes we adopted the {\sf MAG\_BEST} values which are
measured within the elliptical Kron radius (Kron \cite{Kron}) determined
automatically by SExtractor.
To verify our calibration, we also conducted aperture photometry and compared
our aperture magnitudes to published values. We found good agreement 
within our photometric errors (see below). 
For the TF diagram presented in \science, we determined the absolute restframe 
$B$ magnitudes.
To minimize k-corrections, we chose different observed bands as starting point
depending on the redshift of the cluster. 
In the case of \cldrei, the three $R$ band images were coadded to create a 
single frame with a total exposure time of 900s.
While we used the $I$ band image for \clvier, we took the 
1350s $V$ band exposure from the Cluster Deep Field Survey for 
\mseins\ .The latter is the sum of 3 images with the best seeing (FWHM$=$
0\farcs62).
The total apparent magnitudes were corrected for intrinsic extinction 
following the prescription by Tully and Fouque (\cite{TuFo}) with a value of 
$A_B=0.27^m$ for a galaxy seen face-on.
For the conversion from the $B$ band to the other filters we adopted the 
factors: $A_I/A_B=0.45$, $A_R/A_B=0.56$, $A_V/A_B=0.72$.
Depending on the redshift of the cluster, $A_B$ was used for the $V$ image
of \mseins, $A_V$ for the $I$ image of \clvier, and $A_B$ for the $R$ image
of \cldrei.
Galactic extinction was corrected for assuming the values calculated from the
dust maps of Schlegel et al.~(\cite{Schlegel}).
With $E(B-V) = 0.135$ for \cldrei, we used $A_R^g=0.36$. 
With $E(B-V) = 0.041$ for \clvier, we used $A_I^g=0.08$. 
With $E(B-V) = 0.069$ for \mseins, we used $A_V^g=0.23$. 
K-corrections for the transformation between observed filter and restframe
B were calculated according to the SED type 
of the respective galaxy using synthetic model spectra as described in
B\"ohm et al. (\cite{Boehm}).
Finally, the distance modulus for each galaxy was calculated using the
``concordance'' cosmological values ($H_0 = 70$\,km\,s$^{-1}$\,Mpc$^{-1}$,
$\Omega_{\rm m}=0.3$, $\Omega_{\lambda}=0.7$).

Errors due to photon noise were estimated to be very low by SExtractor since
the targeted galaxies are rather bright.
Further errors arise from uncertainties in the calculation of the
k-correction (determination of the T type, see B\"ohm et al. \cite{Boehm}), 
of the internal absorption which depends on the
inclination, and of the Galactic extinction.
We estimate total errors in the magnitudes to be typically 0.10 mag for
galaxies in the field of \mseins, and 0.15 mag in the case of \cldrei\ and
\clvier.
\section{Sample selection}
The selection and observation of cluster spiral candidates was mainly 
determined by two boundary conditions resulting in a heterogeneous sample.
First of all, we utilized previous publications which
had different levels of information on the cluster galaxy population. 

For \cldrei, we used the spectroscopic catalog of
Dressler \& Gunn (\cite{DG92}) and the morphological data from 
HST observations (Smail et al. \cite{Smail}). 
In addition, an unpublished list of Belloni (priv.com.) was
used. In total, 82 galaxies which were i) cluster members, ii) of 
late type and iii) brighter than R$=$ 23 mag were selected of which we 
determined the position angle (PA) from the 
SExtractor photometry on the ground based images. 
These values show a good
agreement with SExtractor measurements of some of the same galaxies 
applied to the HST images which cover only the central cluster parts. 
Then, in the end, 22 appropriate galaxies with 
mean position angles close to the FORS--rotation angles of 
45$^{\circ}$ and -45$^{\circ}$ were selected 
for filling up two different MOS setups. The remaining slitlets 
were filled with 16 other cluster member candidates independent from our 
constraints on Hubble type and PA.

For \clvier, we used the MORPHS data
(Poggianti et al.~ \cite{PSDCB99} and Smail et al.~\cite{Smail})
and ground-based optical-infrared colors of Stanford 
et al.~(\cite{SEDHP02})\footnote{which
were kindly provided well in advance of their actual publication.}.
Since the HST/WFPC2 field covered only a small part of the FORS field and
since there were only 10 spectroscopically verified MORPHS cluster members, 
the primarily selection was done by color. 
By comparing stellar population models
of Bruzual and Charlot (\cite{BC93}, in the 1995 version) with the red 
sequence of ellipticals
determined by Stanford et al.~(\cite{SED98}), 
we applied the following color criteria to select targets with the highest
probability to be cluster members:
spiral candidates had to obey $I\ge19.0, 4.0\le(V-K)\le5.3, 2.2\le(I-K)\le3.15,
1.1\le(J-K)\le1.65$ and $0.3\le(H-K)\le0.87$ (the 10 
spectroscopically verified cluster members lie within these colour ranges).
This resulted in 39 possible
targets, which had to be distributed among two setups. Since the distribution
of the position angles as measured on a 10-minute $I$-band FORS1 pre-image
were rather smooth, the orientation of the two setups were set to differ by
90 degrees ($45^{\circ}$ and $-45^{\circ}$). 
To increase the number of possible targets, we 
produced also a list of galaxies with somewhat redder colors than cited above
as candidates for cluster ellipticals.
Herewith we supplemented our data set of early type cluster galaxies
of another ongoing project.
For the first (second) setup 7 (5) slits had to be filled with
objects with no color information available, of which only 1 turned out to be
a cluster spiral.

The most comprehensive source for creating a candidate list of 
galaxies ($\approx 80$ targets) was available for \mseins . 
In this case we used the catalogue 
of Yee et al. (\cite{YEMAC98}) which contains cluster members and 
their spectral types from the CNOC survey 
({\bf{C}}anadian {\bf{N}}etwork for {\bf{O}}bservational {\bf{C}}osmology).
Thus, galaxies with emission lines (CNOC classes 4 \& 5) were preferentially
assigned to the MOS slitlets of two different setups taking also into account
their respective position angles which we measured on a 5-minute $I$-band
FORS2 pre-image. Together, 14 different spiral candidates, of which 7 were
taken for both setups, were chosen as targets. The remaining slitlets of
the two MOS setups 
were
filled with 3 E$+$A and 8 elliptical cluster candidates and 
serendipetously selected objects from the pre-image.

The main problem for observing an appropriate sample of cluster spiral
galaxy candidates arises from our restriction to the standard MOS unit of FORS 
at the time of the observations. Within one setup this mode provides 19 
slitlets covering the area of the CCD in y--direction.
All of them are individually moveable along the x--axis but have 
a fixed orientation and therefore a fixed slit angle. 
Ideally, slits should be placed along the major 
axis of a galaxy to measure its rotation curve. 
Since this was impossible for the major fraction of our candidates,
we rotated the FORS instrument as to minimize the 
deviation $\delta$ between slit angle and position angle within one setup. 
Unavoidably, the 
deviation $\delta$ was rather large in some cases leading to
geometric distortions of the observed velocity profile that could not be 
fully corrected for. 
\section{MOS--Observations and data reduction}
Observations were carried out in November 1999 (\clvier ), March 2000
(\mseins ) and October 2000 (\cldrei ).
Both FORS instruments were mounted at the Cassegrain focus 
of one of the ESO--VLT unit telescopes, respectively. 
Each instrument had a 2K Tektronix CCD 
with 24$\mu$ pixels and was used in the standard mode. This provides 
a resolution of 0\farcs2/pixel with a total FOV of
6\farcm 8$\times$6\farcm8 in imaging mode and a 
usable FOV of 6\farcm 8$\times$4\farcm 0 for MOS.

In total we gained 125 spectra of 116 objects in the range 
18.0 $<$ R $<$ 23.0 (some objects were observed within two setups). 
For each setup a total exposure time in the order of 8000 sec
(splitted into 3--5 individual exposures)
was chosen to achieve an $S/N \approx 5$ in the emission lines even for the 
faintest galaxies. 

Due to the faintness and the apparent small spatial size of the distant galaxies 
(in the order of 1\arcsec only) particular care has to be taken for
placing an object onto a slit. 
The first order positioning of the slitlets was done on the basis of 
FORS images and the FIMS--software\footnote{FORS Instrument Mask Simulator, 
see http://www.eso.org}. After taking a through--slit aquisition 
frame the final positioning of the slitlets on the targets was 
within an accuracy of 0\farcs1.   
 
We used a slitwidth of 1\arcsec and grism GRIS 600R\footnote{Grism 600R$+$14
with order separation filter GG435 at FORS1, Grism 600R$+$24
with order separation filter GG435$+$81 at FORS2}.
This provides a resolution of $\lambda/\Delta\lambda=$1230 
and a dispersion of 45\AA/mm corresponding to a sampling of 1.08 \AA/pixel.
Each individual spectrum covers typically 2000 \AA ~within a wavelength 
range between 4400--8200 \AA ~depending on the slit position within the 
FOV. 
\begin{figure}
\centerline{\hbox{
\psfig{figure=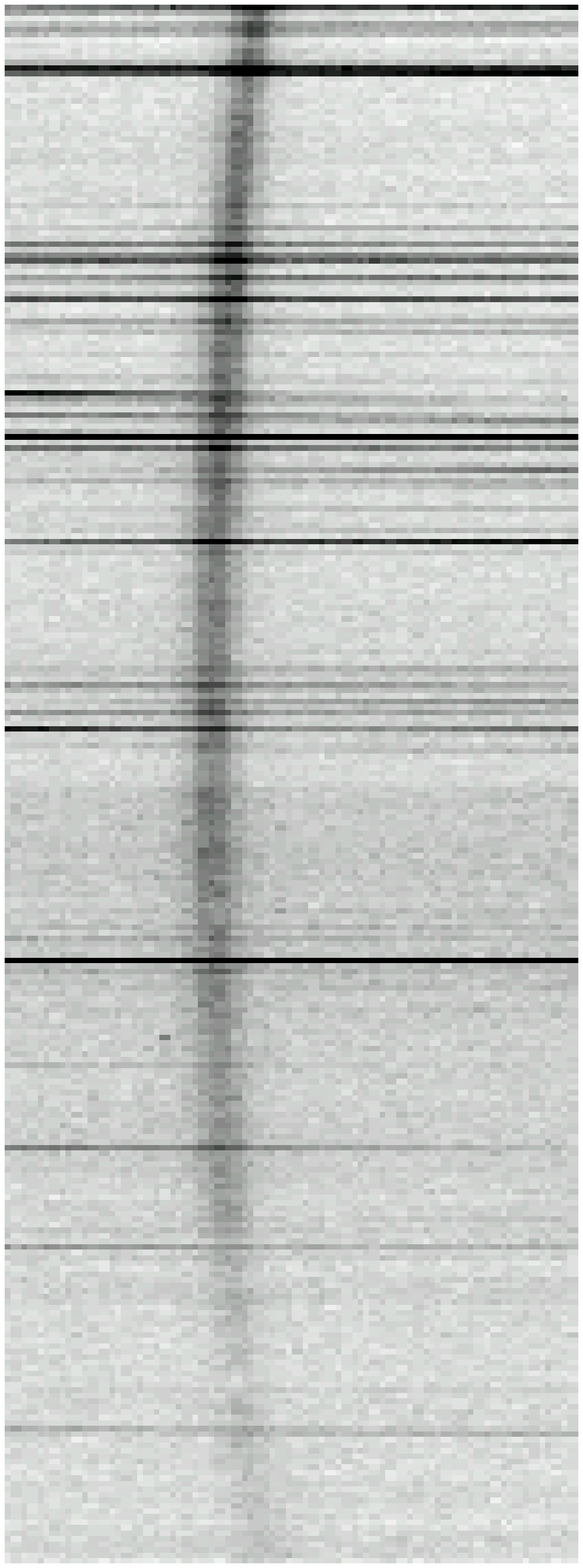,width=8cm,angle=-90,clip=t}
}}
\centerline{\hbox{
\psfig{figure=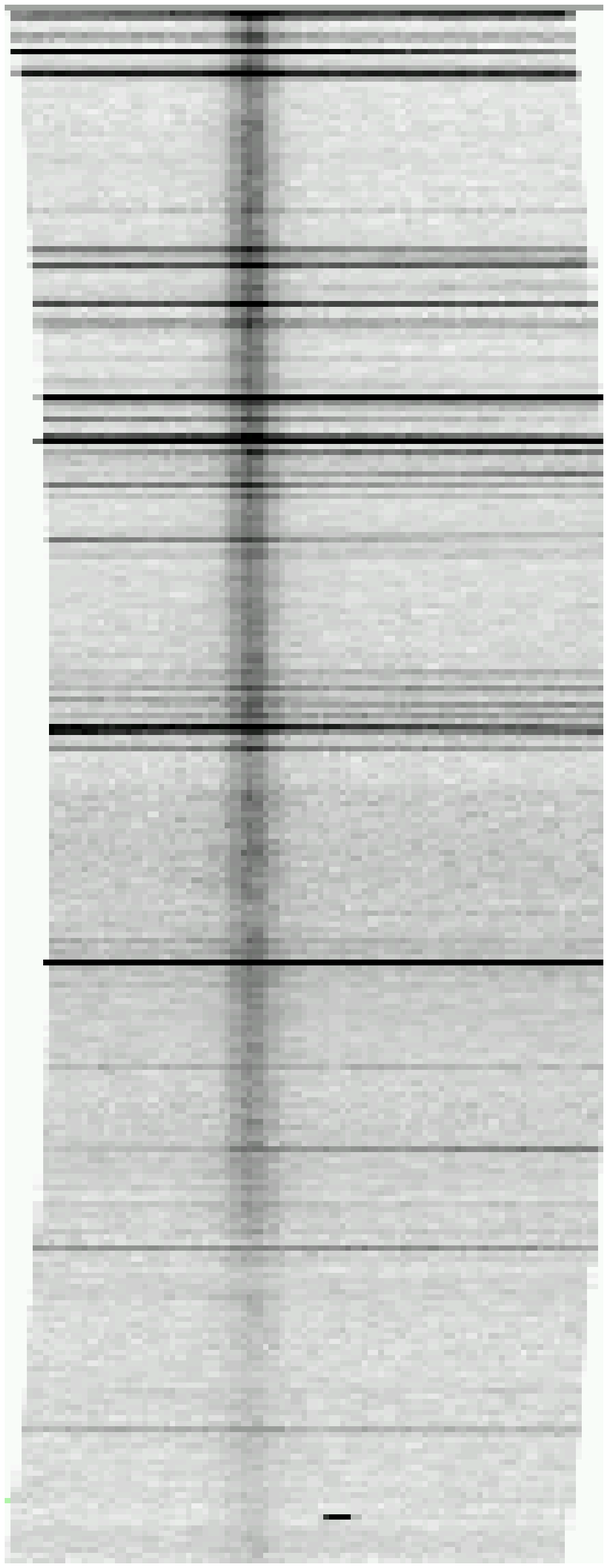,width=8cm,angle=-90,clip=t}
}}
\centerline{\hbox{
\psfig{figure=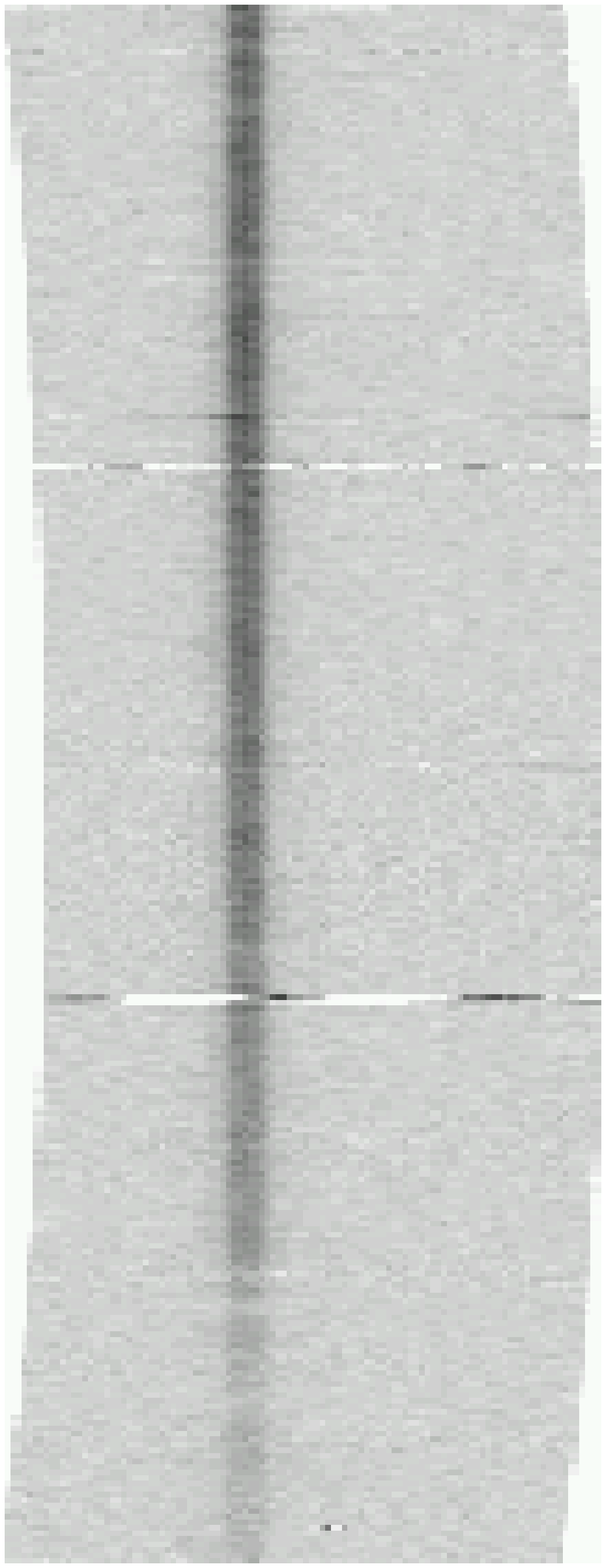,width=8cm,angle=-90,clip=t}
}}
\caption {
Example for the correction of the  
curvature of the spectra in y--direction. 
{\bf{Top}}. The raw but cosmic corrected two dimensional spectrum of the 
galaxy A18 (slit no.~18 of setup A) of \mseins\ . 
{\bf{Center}}. The same 
spectrum after applying the rectification. 
{\bf{Bottom}}. The fully reduced and wavelength calibrated
spectrum.  
A18 is an early--type galaxy and a member of \mseins\  ($z=0.3111$).
Please note that the figures are expanded in y--direction for a
better illustration of the curvature.}
\end{figure}
All data reduction steps were done with MIDAS 
or with own procedures which were implemented into MIDAS.

Since the ESO--Paranal staff provides calibration frames 
(as bias, flatfields or standards) 
of VLT observations in general by following a standard calibration plan 
we have checked at first all calibration frames for compatibility 
with their corresponding science frames (e.g.~same read--out--mode, 
sampling, gain etc.). 
Furthermore, properties of sets of calibration frames 
contemporary gained with the science frames during one night have been 
principally compared with other sets e.g.~before averaging.
In some cases science frames of one setup were collected during 
several nights. Such sub--setups and their corresponding 
calibration frames were always treated separately at first to check at
which state of the data reduction they could be combined.
\begin{figure*}
\centerline{\hbox{
\psfig{figure=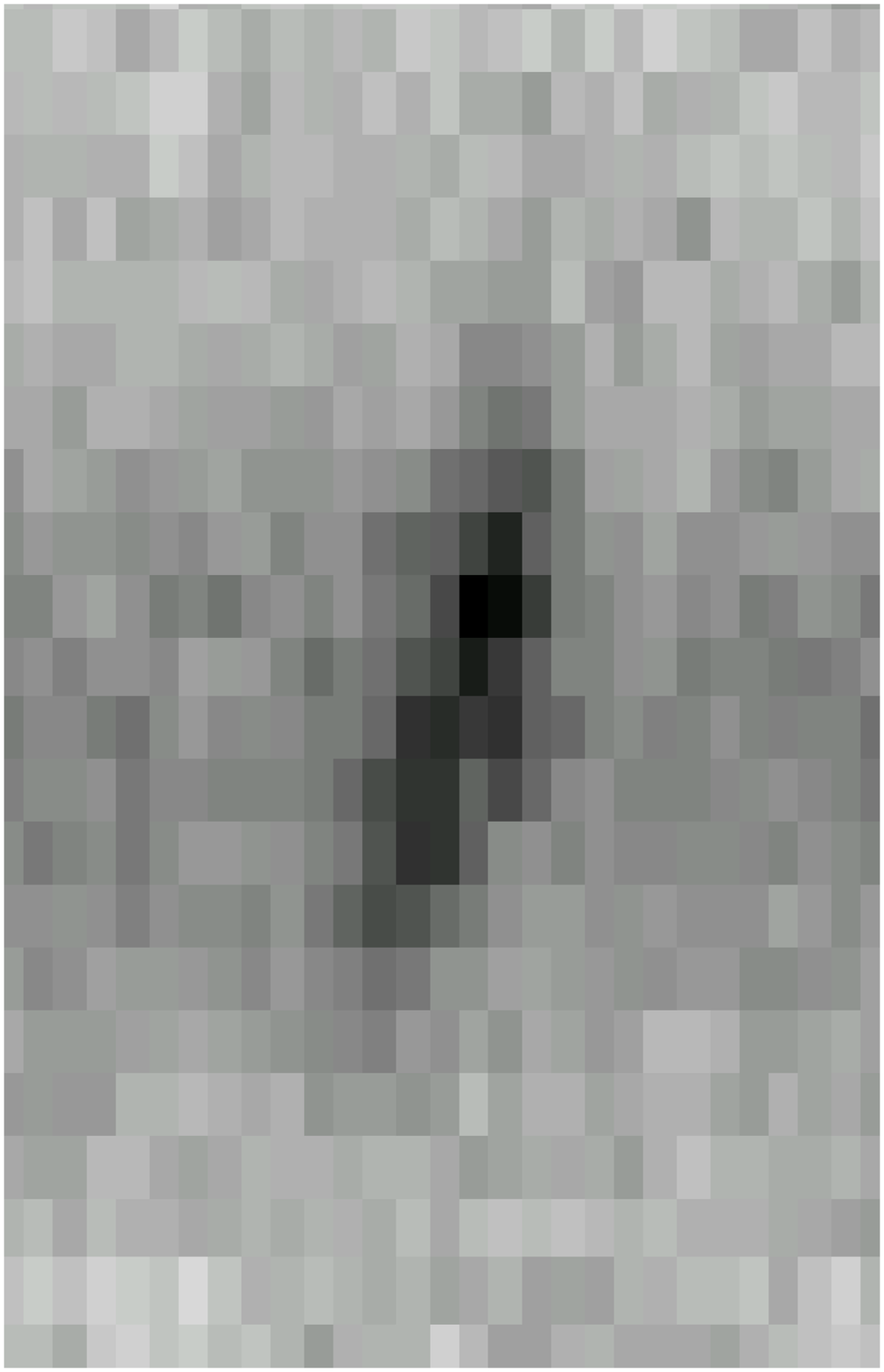,width=3.5cm,clip=t}
\psfig{figure=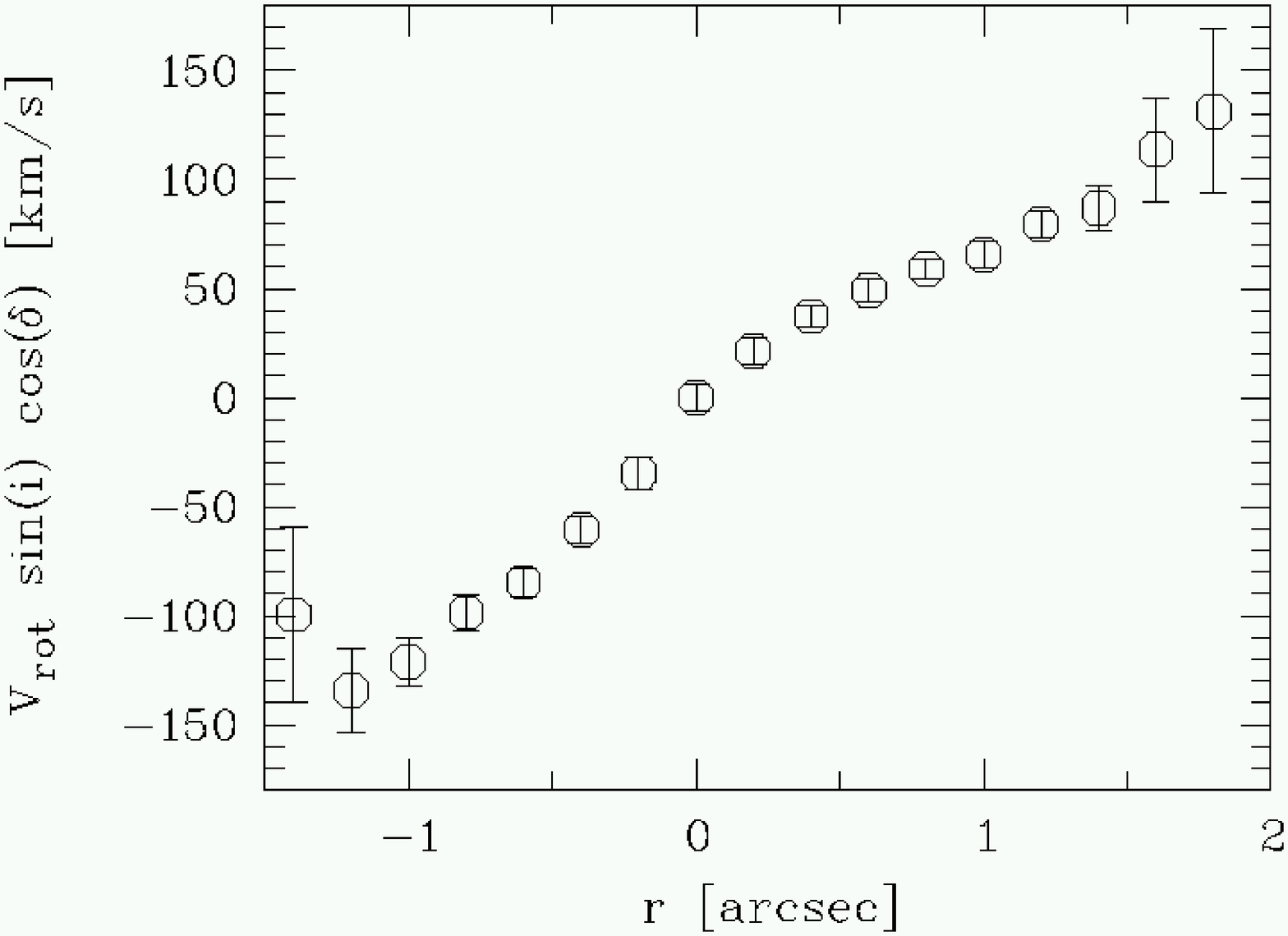,width=8.5cm,angle=-90,clip=t}
\psfig{figure=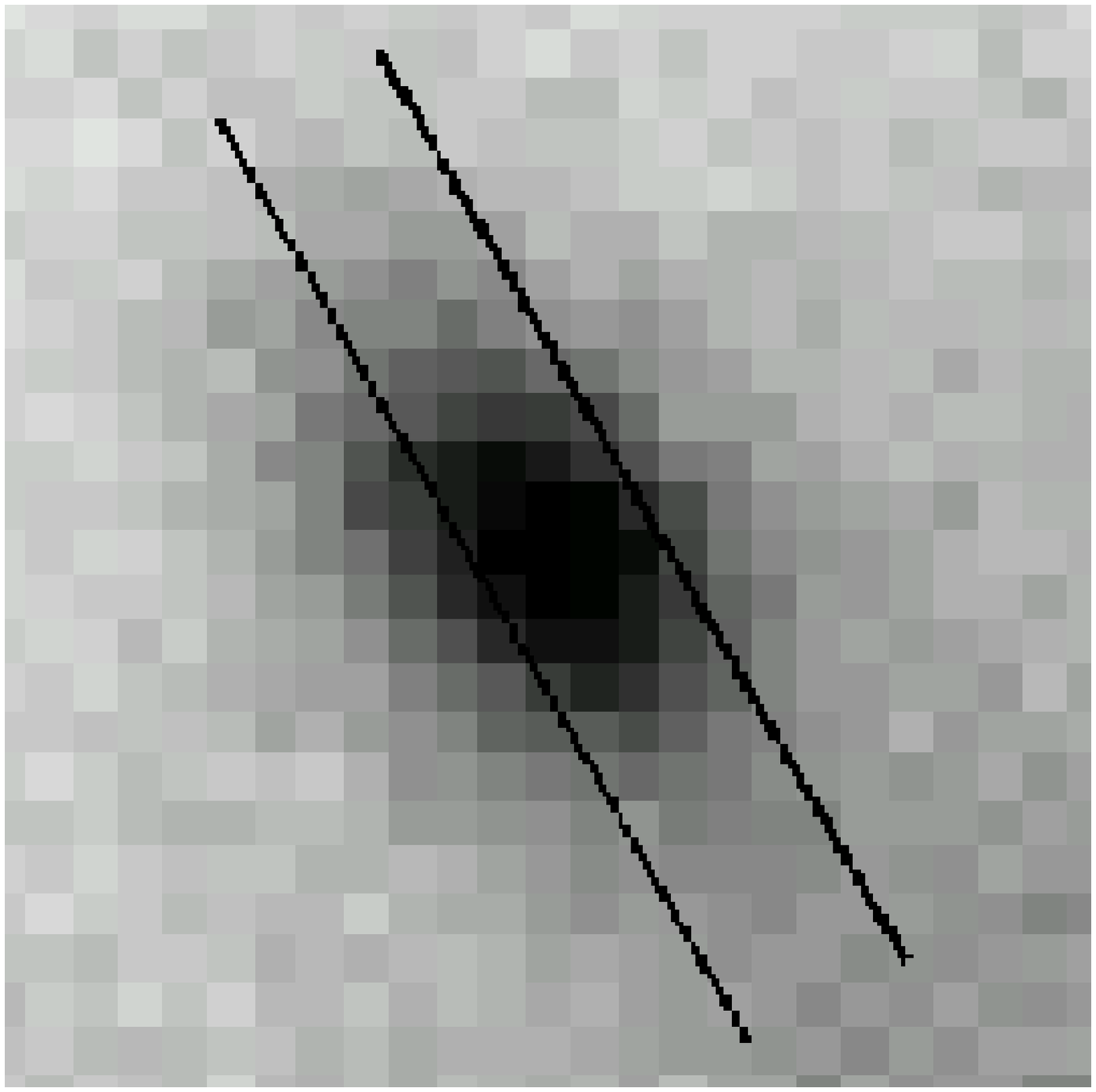,width=5.5cm,clip=t}
}}
\caption {Illustration of our observed data by means of one galaxy from Setup B
of the cluster \mseins\  (ID$=$B5). {\bf{Left.}}
The spatially resolved $H_{\beta}$--emission line showing clearly the Doppler
shifts due to the rotation of the galaxy (for a better illustration 
the image was expanded in the spatial direction).  
{\bf{Center.}} The corresponding measured rotation curve. 
{\bf{Right.}} Image of the galaxy extracted from the FORS 
frame. The galaxy with $R=20.70$ has an inclination of
60$^{\circ}$ and is a cluster member ($z=0.3039$). The slit position 
is marked.} 
\end{figure*}
Firstly, all science and flatfield frames were bias corrected.
We created master bias frames for each observing 
night by the median of typically 20 single bias frames which have been
set to the same level before. This was possible since in all cases 
the mean bias level 
changed only in the order of $\ll 1$\% even between different nights and all 
frames showed little but equal two dimensional patterns. 
Then, each master bias was scaled to the overscan level of 
the science frame and subtracted. 
Hereafter, a cosmic ray correction on the two dimensional 
MOS frames has been applied by using the standard MIDAS tool FILTER/COSMIC.
Within this routine cosmic ray events are detected by the comparison of the raw
frame with a median filtered version of that image. Pixels with an intensity
greater than a defined threshold as compared to the local median are replaced
by this median value. To avoid misdetections, in particular within the emission
lines, we furthermore compared the individual exposures of
a given setup.

Now, two dimensional subframes corresponding to individual slit spectra 
were extracted from each MOS frame. 
Similarly, appropriate subframes (with same dimensions and CCD--pixel 
coordinates) were extracted from the corresponding two dimensional 
flatfield-- and calibration images. Flatfields were gained with
through--slit--exposures. For each setup two versions with
different exposure times were taken to warrant for correct 
illumination of the upper and lower part of the CCD, respectively.
This was necessary to avoid diffuse reflection from the gaps between the
slitlets. To perform the correction of the science spectra 
for pixel--to--pixel variations and the CCD-response over the dispersion 
direction masterflats were created from typically 3--5 
corresponding single flatfields by normalization and averaging.\\[2mm]

One of the critical steps was the correction for an instrument based 
curvature of the frames in y--direction (see Fig.1 for an illustration). 
This distortion changes gradually from a positive curvature at the top of the
CCD to a 
negative curvature at the bottom of the CCD with a maximal misalignment in the
order of five pixel (in y) at the edges of the uppermost and lowest
spectrum. To correct for, each column of a certain slit image 
has to be shifted with subpixel accuracy. To get values with a resolution of 
0.1 pixel we applied a fit to the curvature.
We checked that the flux of the science
frames was conserved within a few percent. 
The same procedure was applied to the 
calibration spectra. 
We point out that the rectification for corresponding science/calibration
spectra has to be achieved
with exactly the same parameters. Furthermore, this reduction step has to be 
done after the flatfield correction since there are small but non-commutative 
changes of pixel values which would affect the flatfield correction 
if doing it in the reverse order.

To subtract the night sky emission typically 30--40 rows \newline 
($\approx$ 6\arcsec--8\arcsec) on both sides of
the galaxy emission were selected interactively 
before each column of a certain spectrum was fitted. 
Only few galaxy spectra were located very close to the edges of the 
respective slitlets. In those cases the definition of the sky 
region was restricted to
one side of the galaxy spectrum, only.

Distortions in wavelength direction, i.e.~the curvature of arc--lines 
were corrected during the 
standard wavelength calibration. Calibration spectra have been taken with 
Ne, He, and Ar lamps and a two--dimensional dispersion relation 
(rms $\approx 0.04 \AA$) was calculated for each spectrum 
by polynomial fits.
This dispersion relation was then applied to the corresponding 
science spectrum.

For the summation of single exposures of the same slitlet, 
the center of the galaxy emission along the spatial axis was 
determined by a Gaussian fit for
each exposure. Only for a very limited number we measured an offset 
in the order of 
1--2 pixel making it necessary to shift along the spatial axis before summation.
In the case of seeing variations larger than 25\% between individual frames 
a weighted addition has been applied.
\section{Modelling of rotation curves}
For the derivation of rotational velocities of the galaxies 
as a function of radius we followed exactly the same procedure as
described in detail in Ziegler et al. (\cite{Ziegler2}) and 
in B\"ohm et.al (\cite{Boehm}).
Both publications present the analysis of the internal kinematics of 
spiral galaxies at intermediate redshift taken from the FORS Deep Field
Survey (e.g.~Heidt et al.~(\cite{JH}). 
Since these objects were observed with the same instrumental setups as
the ones presented here we only summarize the main properties of this part 
of the analysis.
\subsection{Observed rotation curve RC$_{obs}$}
Rotation velocities were determined by an analysis of the
[OII] 3727, H$_\beta$ or [OIII] 5007 emission lines (see Fig.2 for 
an illustration of our data).
Any blue-- and redshift in wavelength within an emission line 
due to the internal kinematics of a galaxy has to be measured with 
reference to the center of that line. 
Therefore, we firstly fitted a Gaussian profile to the selected 
emission line to define its center along the spatial axis within an accuracy 
of 0\farcs1. This was achieved by averaging 100 columns centered on that 
emission line to construct a one--dimensional intensity profile along the
spatial axis.

To determine the wavelength shifts along the spatial axis the emission line 
was fitted row by row 
with Gaussian profiles after three neighbouring rows were averaged 
before each fit to increase the S/N ratio. 
Due to the resolution of 0\farcs2 arcsec/pixel this "boxcar"--filter corresponds 
to 0\farcs6 (only in the few cases of very weak lines we used a boxcar of 
1\arcsec). Wavelength shifts relative to the center of the line 
were now transformed into velocity shifts under consideration of 
a cosmological correction of $(1+z)^{-1}$ to gain an observed 
rotation curve RC$_{obs}$. 

Any velocity value measured on the basis of RC$_{obs}$ 
will lead to an underestimation of the true intrinsic maximum galaxy 
rotation velocity.
This is due to the fact that the visible disk size of spirals at 
intermediate redshifts and the slit width of 1 arcsec
are of comparable sizes and, therefore, an integrated spectrum 
covers a substantial fraction of
the two--dimensional intrinsic velocity field of a galaxy. At $z=0.5$ -- 
that is the distance of \clvier\ for example -- a typical 
scalelength of 3kpc corresponds to $\approx$0.5 arcsec.  
Hence, this "blurring" effect has to be taken into account in particular 
for the determination of the maximum rotation velocity V$_{max}$. 
The problem was solved by generating a synthetic rotation 
curve RC$_{syn}$ which was compared with RC$_{obs}$.

\subsection{Synthetic rotation curve RC$_{syn}$}
To simulate a rotation curve we assumed an intrinsic rotation law. 
A simple shape was used with a 
linear rise of V$_{rot}$ at small radii turning over at a characteristic radius 
$r_{0}$ into a flat part of V$_{rot}=const=$V$_{max}$ as it can be expected 
due to the influence of a Dark Matter Halo. 

Based on RC$_{syn}$ we generated the two--dimensional velocity 
field for an individual galaxy as it would appear under consideration of 
the observed position angle, inclination and disk scale length as determined 
by a fit to the respective two--dimensional surface brightness profile 
and the seeing (FWHM) during spectroscopy. 
Simulating the slit spectroscopy we then integrate over the velocity field
within a stripe of 1\arcsec taking also into account the mismatch angle
$\delta$ to find V$_{max}$. This fit was repeated varying the values of 
$i$, $\delta$ and $r_{d}$ within the errors. For a detailed 
discussion of a $\chi^2$-fitting procedure based on the errors from the 
RC extraction and a discussion of the influence of different RC$_{syn}$ 
shapes on V$_{max}$ see also B\"ohm et al. (\cite{Boehm}). 

In 8 cases more than one emission line was present within a spectrum.
Rotation curves have been measured from each line and compared. 
They were consistent within the errors in 6 cases.
The other two had too low S/N.
The tabulated values for $V_{max}$ have been derived 
from the curve with the largest covered radius and the highest S/N
ratio. 
\subsection{Disturbed kinematics of cluster spirals}
Only galaxies showing a rotation curve that is 
rising in the inner region and then turning into a flat part 
are useful for the construction of a TF--diagram. 
In such cases the maximum
rotation velocity can be used as an indicator for the total dynamical mass of a
galaxy and would therefore reflect the influence of a dark matter halo on the
internal galaxy kinematics. 
In \science\ we presented a TF--diagram in which 13 cluster 
spirals of \cldrei\ , \clvier\ , and \mseins\  as well as seven
field spiral galaxies are shown in comparison to the distribution of the FORS Deep
Field galaxies from B\"ohm et al. (\cite{Boehm}). We point out that we only used those
galaxies for the diagram where we were able to measure V$_{max}$ in the sense
mentioned above. Galaxies with disturbed kinematics (either intrinsic or
geometric) do not enter the TF--diagram. 
\section{The data}
Within this section we present data tables (Table 2--10) on individual galaxies, 
finding charts (Fig.3) of the six MOS--setups, and position velocity diagrams
of galaxies (Fig.4--6) in the field of the clusters.

The finding charts show the full 
FORS 6\farcm 8 $\times$ 6\farcm8 --FOV
in standard imaging mode.
All observed primary MOS--targets are marked by circles and 
labeled by their slit numbers.  These numbers correspond to the identifier
(ID) given in the data tables. Table 2--4 contain a complete list of 
all observed objects (within all setups and all slitlets) with
positions, redshifts and magnitudes.
The positions were measured on our FORS images and are 
not astrometric corrected.

In Table 5 specific information is given on only those spiral galaxies for
which we derived V$_{max}$.
The columns of that table have the following meaning, respectively:
\begin{enumerate}

\item cluster -- cluster setup of MOS--spectroscopy.

\item ID --- Identification number of galaxies corresponding to
the setup (A or B) and the slit number. 
 
\item mem --- shows cluster membership of a galaxy. 

\item T --- type of SED in the de Vaucouleurs scheme.  
$T=1$ corresponds to Hubble type Sa, $T=3$ to Sb, $T=5$ to Sc, and 
$T=8$ to Sdm. Classification criteria as described in 
B\"ohm et al.~(\cite{Boehm}).

\item incl.--- Disk inclination derived by minimizing $\chi^2$ of an
exponential profile fit to the galaxy image extracted from our FORS frames.

\item $\delta$ --- angle of misalignment between the slit and the
apparent major axis of a galaxy. 

\item r$_{d}$ --- apparent disk scale length of a galaxy (in arcsec) 
as derived from ground based data.

\item V$_{max}$ --- Intrinsic maximum rotation velocity of the galaxy 
(see sect.~5 for details). The Error of the intrinsic maximum rotation 
velocity was estimated via $\chi^2$-fits of the synthetic RC to the
observed RC (see B\"ohm et al.~\cite{Boehm}). 
\item EW [OII] --- rest frame equivalent width of the [OII] 
emission line as derived from our spectra.

\item M${_B}$ --- Absolute $B$-band magnitude.
\end{enumerate}

In total we 
gained spectra of 116 objects within the magnitude 
range 18.0 $<$ R $<$ 23.0. We were able to measure 
redshifts and to determine spectral types for 89 galaxies. 
72 turned out to be late type galaxies. 
From the 50 galaxies which were found to be cluster members 35 are late type
galaxies. Finally, for 32 galaxies (15 cluster members)
we were able to obtain position velocity diagrams (Fig.4--6). 
The corresponding 
measured values (V$_{rot}$ versus position) are presented in Tab.~6--10.
13 cluster galaxies exhibit a rotation
curve of the universal form rising in the inner region and passing over into a
flat part. In those cases (plus 7 field galaxies) 
V$_{max}$ could be determined.
The other members have peculiar kinematics or too low
signal--to--noise.
\begin{acknowledgements}
We thank ESO and 
the Paranal staff for efficient support. 
We also thank the PI of the FORS project, Prof. I. Appenzeller (Heidelberg), and
Prof. K. J. Fricke (G\"ottingen) for providing guaranteed time for our project.
Furthermore we want to thank the anonymous referee for his helpful remarks.
This work has been supported by the Volkswagen Foundation (I/76\,520)
and the Deutsche Forschungsgemeinschaft (Fr 325/46--1 and SFB 439).
\end{acknowledgements}

\begin{figure*}
\centerline{\hbox{
\psfig{figure=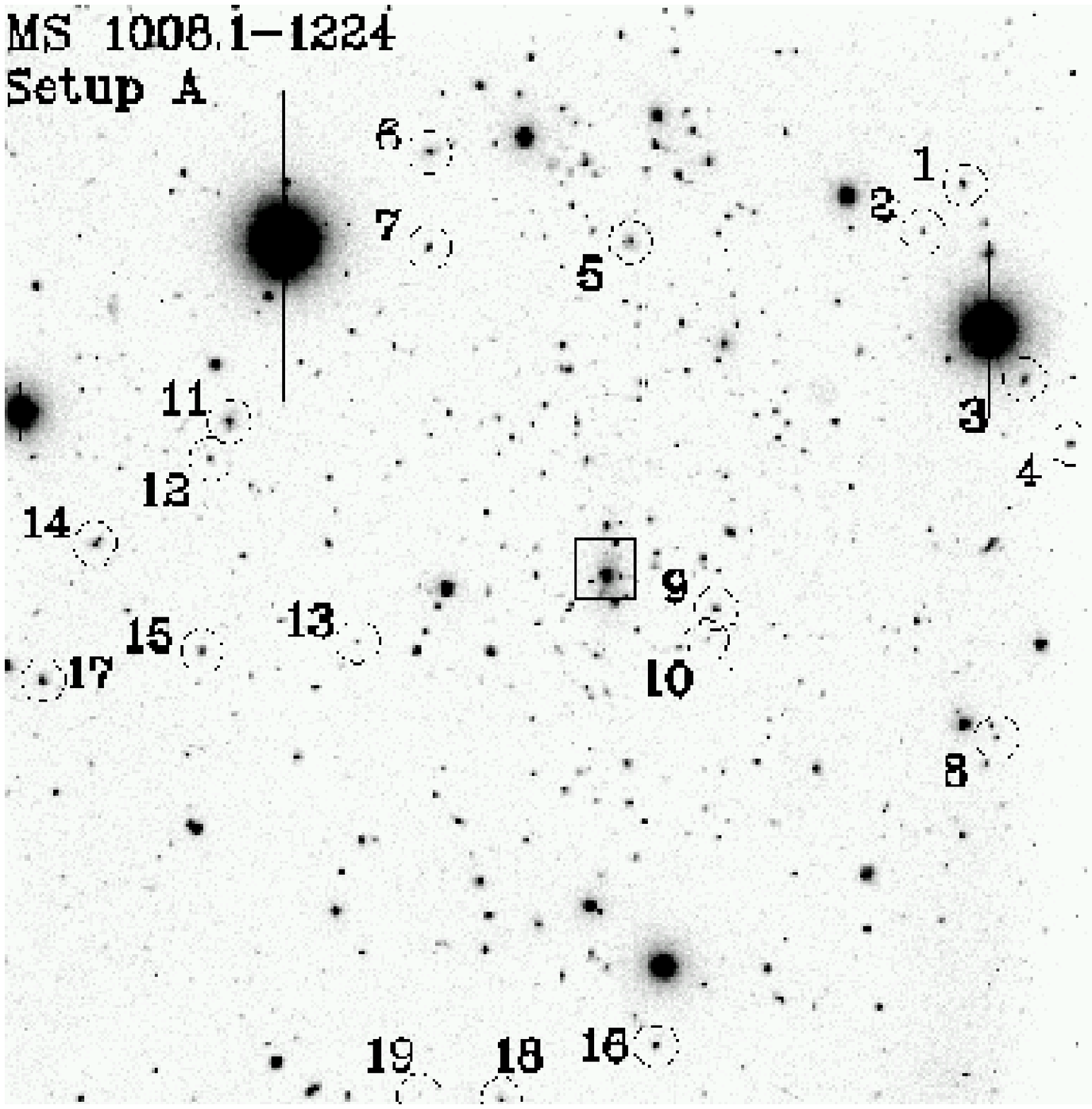,width=7cm,height=7cm,clip=t}
\psfig{figure=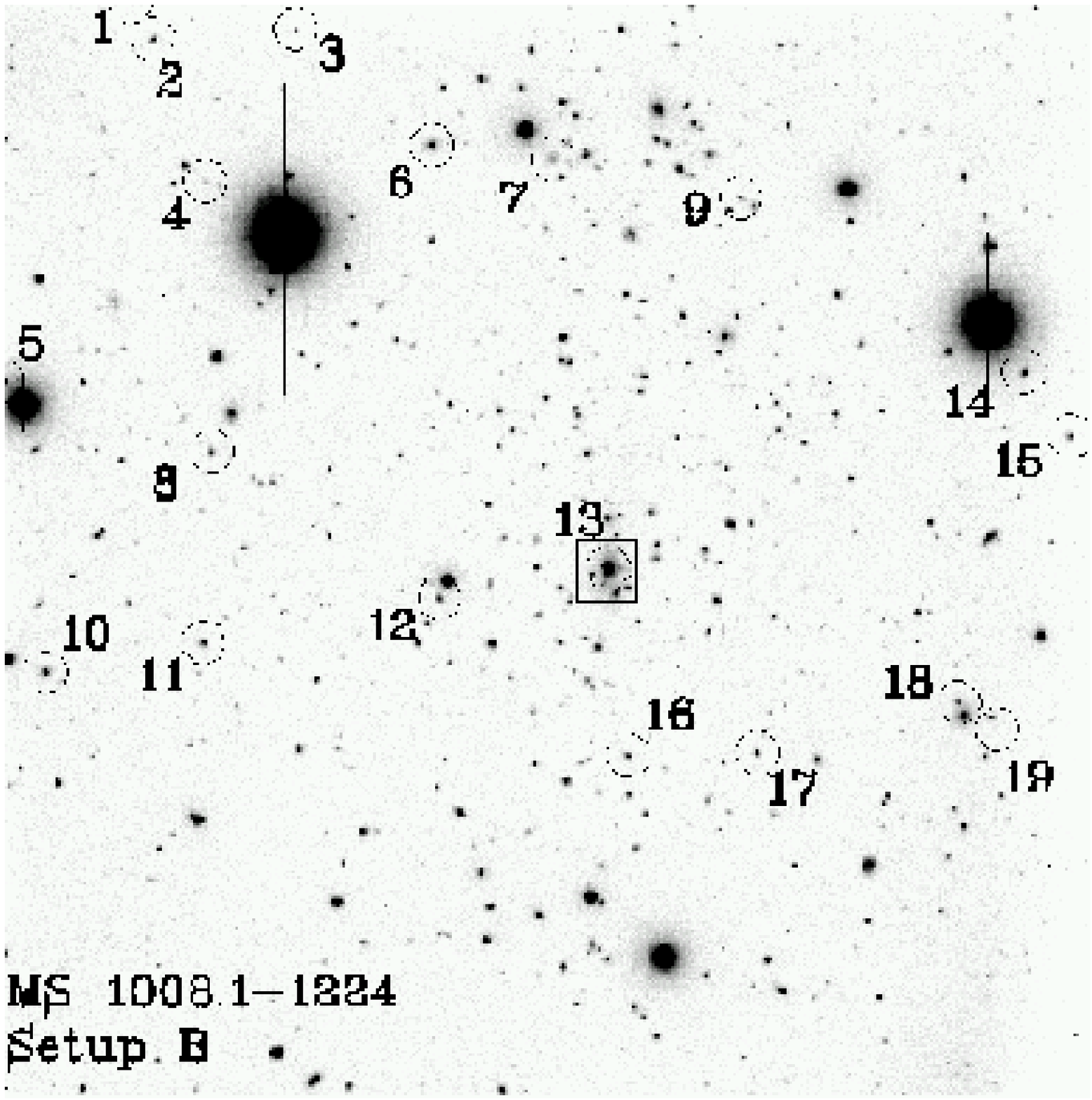,width=7cm,height=7cm,clip=t}
}}
\end{figure*}
\begin{figure*}
\centerline{\hbox{
\psfig{figure=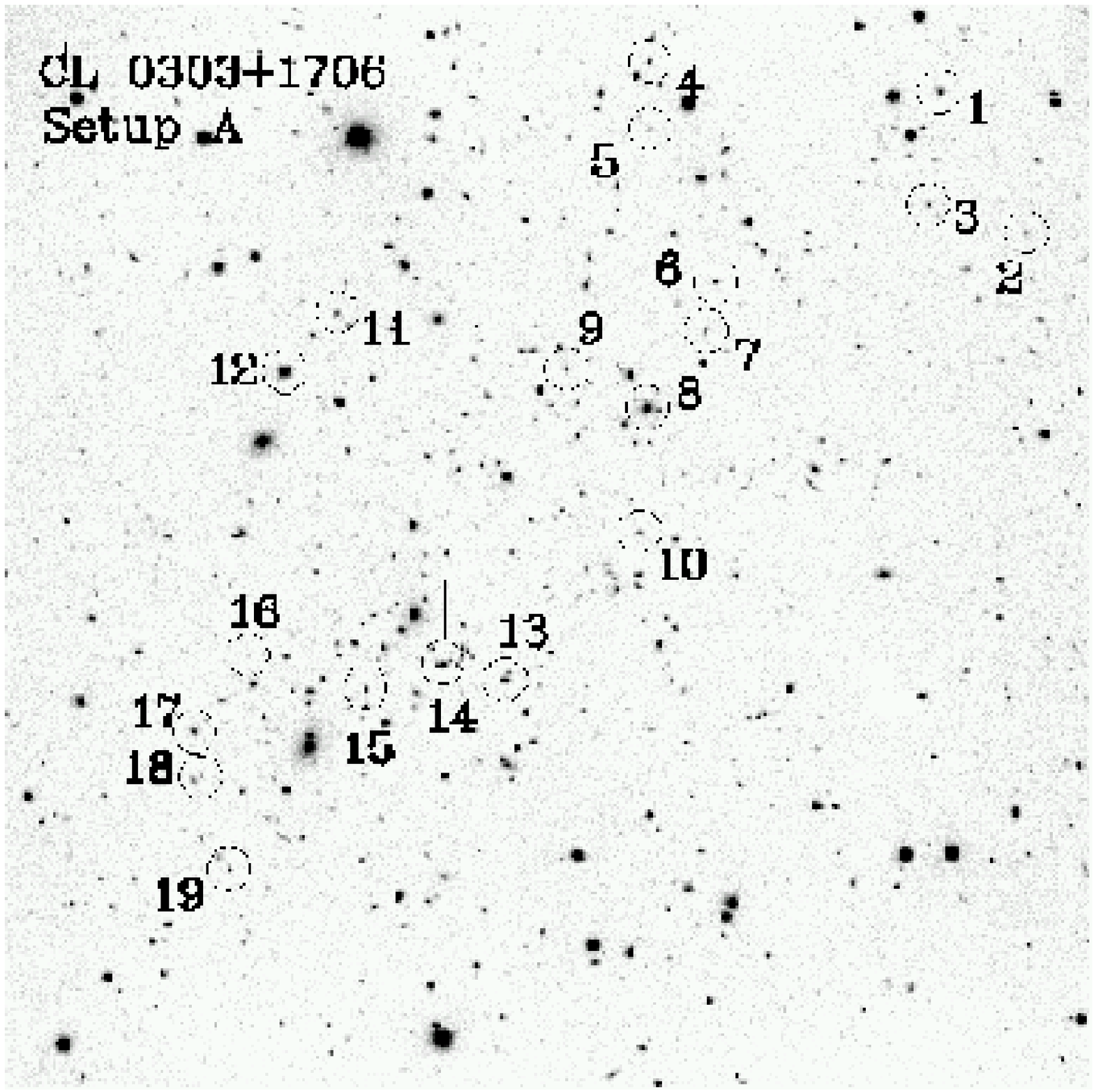,width=7cm,height=7cm,clip=t}
\psfig{figure=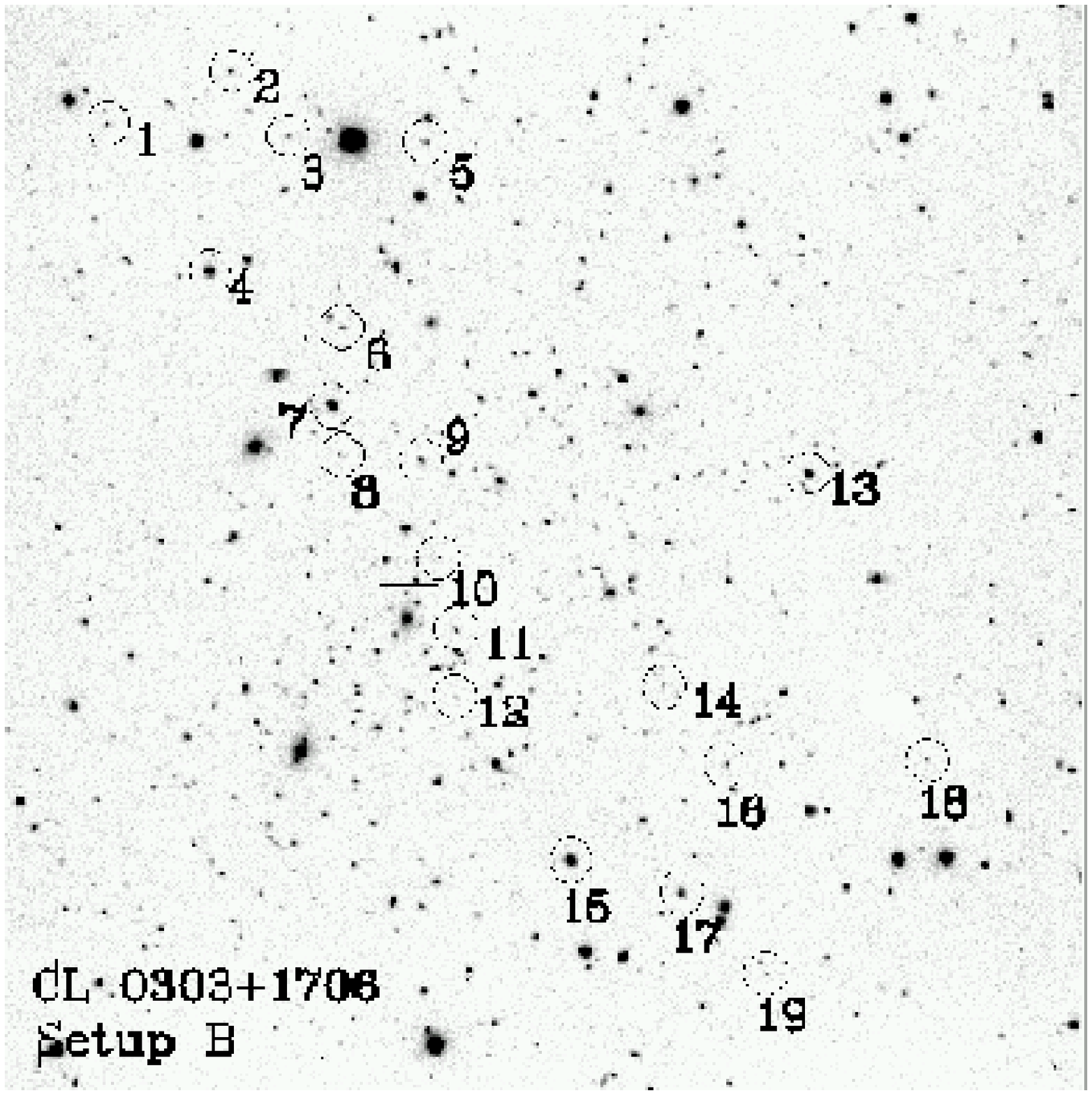,width=7cm,height=7cm,clip=t}
}}
\end{figure*}
\begin{figure*}
\centerline{\hbox{
\psfig{figure=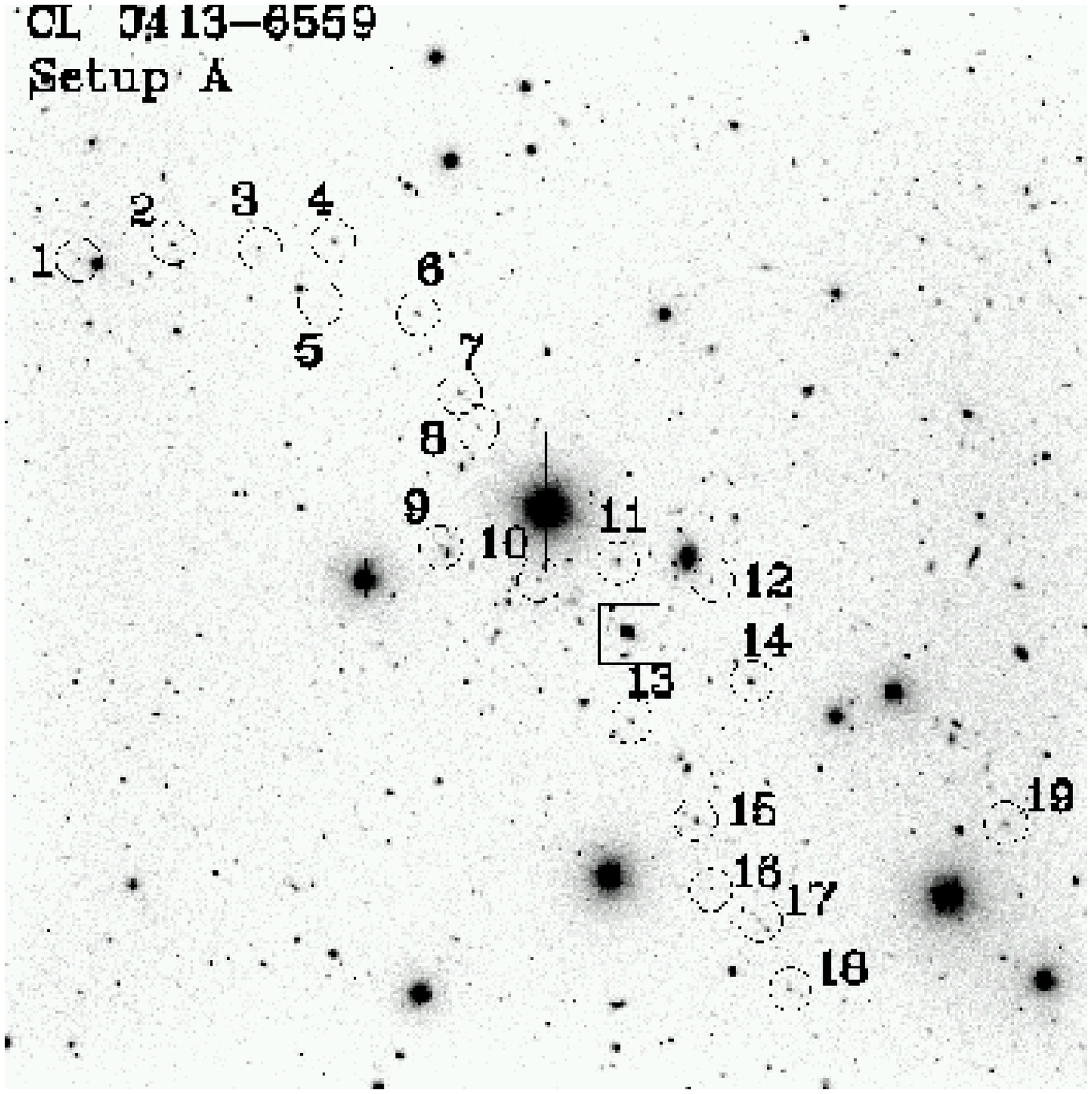,width=7cm,height=7cm,clip=t}
\psfig{figure=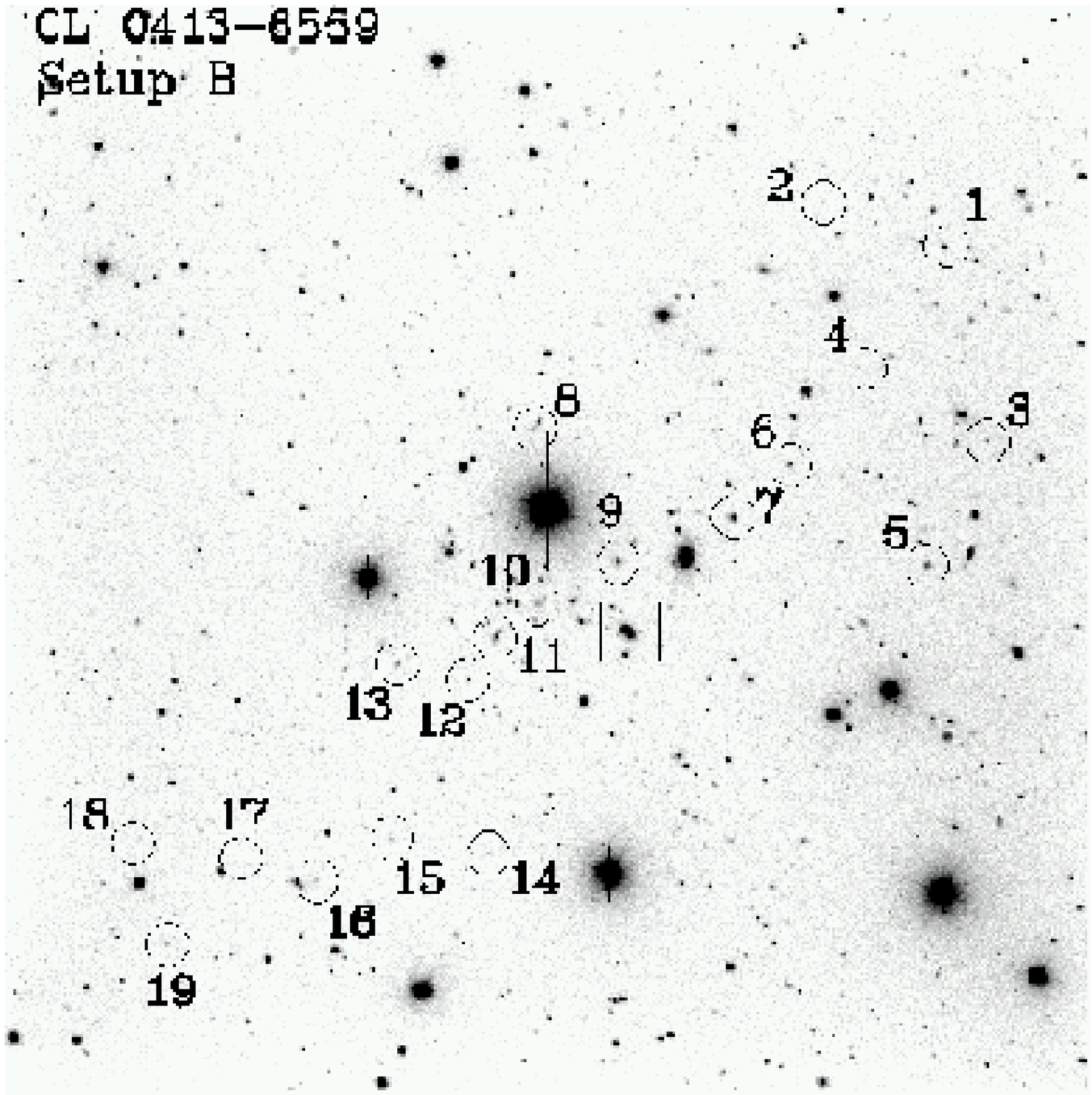,width=7cm,height=7cm,clip=t}
}}
\caption {Finding charts of all MOS setups. {\bf{Top}}. \mseins .
{\bf{Center}}. \cldrei . {\bf{Bottom}}. \clvier .   
In all cases we show the full FORS 6\farcm 8 $\times$ 6\farcm8 --FOV.
All primary MOS--targets are marked by circles and 
labeled by their slit numbers. The central cluster galaxy is marked by a 
square (north is up, east to the left).}
\end{figure*}
\begin{table*}
\centering
\caption{MOS targets in the field of \mseins\ observed with Setup A (PA$=+55^{\circ}$) and 
Setup B (PA$=-58^{\circ}$). 
There are 7 objects which have been observed in both setups: A3$=$B14, A4$=$B15, A6$=$B6, A8$=$B19, A12$=$B8,
A15$=$B11 and A17$=$B10. For identification of galaxies see finding chart in 
Fig.3.}
\begin{tabular}{ccccc|ccccc}
\hline
ID     &RA	 &DE		    & z      &V     &ID    &RA	 &DE	       &z   &V\\    
       &[h m s]  &[${\circ}$ ' '']  &	 &[mag] &     &   [h m s]   &[${\circ}$ ' ''] &       &[mag]	    \\     
\hline
A1    &10 10 23.3&-12 37 27 & 0.3074 &      & B1  &10 10 44.7&-12 36 22 & 0.3222 &20.31\\
A2    &10 10 24.4&-12 37 46 & 0.3167 &      & B2  &10 10 44.1&-12 36 36 & 0.1673 &20.58\\
A3    &10 10 21.8&-12 38 41 &	     &      & B3  &10 10 40.5&-12 36 32 & 0.4256 &21.46\\
A4    &10 10 20.6&-12 39 05 & 0.2960 &20.31 & B4  &10 10 42.8&-12 37 30 &	 &\\
A5    &10 10 31.9&-12 37 49 & 0.3090 &19.84 & B5  &10 10 48.0&-12 38 44 & 0.3039 &\\
A6    &10 10 37.0&-12 37 15 & 0.3526 &19.98 & B6  &10 10 37.0&-12 37 15 & 0.3516 &19.98\\
A7    &10 10 37.0&-12 37 51 & 0.3101 &21.45 & B7  &10 10 33.9&-12 37 21 & 0.3623 &20.63\\
A8    &10 10 22.5&-12 40 54 & 0.3137 &21.13 & B8  &10 10 42.6&-12 39 10 & 0.3088 &20.76\\
A9    &10 10 29.7&-12 40 06 & 0.3044 &20.76 & B9  &10 10 29.1&-12 37 36 & 0.2960 &21.02\\
A10   &10 10 29.9&-12 40 18 & 0.2998 &21.52 & B10 &10 10 46.9&-12 40 33 & 0.3024 &19.78\\
A11   &10 10 42.1&-12 38 56 & 0.3089 &19.64 & B11 &10 10 42.8&-12 40 22 & 0.3219 &20.14\\
A12   &10 10 42.6&-12 39 10 & 0.3090 &20.76 & B12 &10 10 36.8&-12 40 06 & 0.2980 &20.78\\
A13   &10 10 38.9&-12 40 18 & 0.3047 &21.61 & B13 &10 10 32.5&-12 39 54 & 0.3060 &19.54\\
A14   &10 10 45.6&-12 39 42 & 0.3066 &20.58 & B14 &10 10 21.8&-12 38 41 & 0.3100 &\\
A15   &10 10 42.8&-12 40 22 & 0.3219 &20.14 & B15 &10 10 20.6&-12 39 05 & 0.2957 &20.31\\
A16   &10 10 31.2&-12 42 49 & 0.3094 &20.82 & B16 &10 10 32.0&-12 41 04 & 0.4644 &21.62\\
A17   &10 10 46.9&-12 40 33 & 0.3022 &19.78 & B17 &10 10 28.6&-12 41 03 & 0.3116 &20.63\\
A18   &10 10 35.2&-12 43 09 & 0.3111 &20.70 & B18 &10 10 23.5&-12 40 45 &	 &20.80\\
A19   &10 10 37.3&-12 43 08 &	     &20.71 & B19 &10 10 22.5&-12 40 54 &	 &21.13\\
      & 	 &	    &	     &      & B7b &10 10 33.7&-12 37 28 & 0.2983 &\\
      & 	 &	    &	     &      & B8b &10 10 42.8&-12 39 06 & 0.3040 &\\
\hline
\end{tabular}
\end{table*}
\begin{table*}
\centering
\caption{MOS targets in the field of \cldrei\ observed with Setup A (PA$=+45^{\circ}$) and 
Setup B (PA$=-45^{\circ}$). A11$=$B6.}
\begin{tabular}{ccccc|ccccc}
\hline
ID    &RA        &DE        &z	 &R	&ID    &RA	  &DE	      &z	   &R\\   
      &[h m s]   &[${\circ}$ ' '']&     &[mag] &      &[h m s]   &[${\circ}$ ' '']&      &[mag] \\	 
\hline  		       
A1    &03 06 05.3&17 22 04& 0.5233 &21.03 &B1	 &03 06 26.8&17 21 53& 0.4232 &22.13 \\
A2    &03 06 03.0&17 21 12&	   &21.59 &B2	 &03 06 23.6&17 22 13& 0.4203 &21.01 \\
A3    &03 06 05.6&17 21 22& 0.4231 &21.18 &B3	 &03 06 22.1&17 21 49&     &21.23 \\
A4    &03 06 12.9&17 22 16&	   &21.40 &B4	 &03 06 24.2&17 20 59& 0.4237 &19.44 \\
A5    &03 06 12.9&17 21 50& 0.4174 &21.83 &B5	 &03 06 18.5&17 21 47&     &20.97 \\
A6    &03 06 11.2&17 20 54& 0.4043 &21.67 &B6	 &03 06 21.1&17 20 42& 0.4142 &21.53 \\
A7    &03 06 11.4&17 20 35& 0.4190 &20.81 &B7	 &03 06 21.0&17 20 09& 0.3597 &19.25 \\
A8    &03 06 12.9&17 20 06& 0.4207 &19.30 &B8	 &03 06 20.7&17 19 51&     &22.75 \\
A9    &03 06 15.1&17 20 21& 0.4039 &21.11 &B9	 &03 06 18.7&17 19 49& 0.4174 &21.43 \\
A10   &03 06 13.1&17 19 20& 0.6413 &21.77 &B10   &03 06 18.2&17 19 12&     &21.30 \\
A11   &03 06 21.1&17 20 42&	   &21.35 &B11   &03 06 17.8&17 18 45& 0.4187 &21.42 \\
A12   &03 06 22.4&17 20 20& 0.1505 &19.37 &B12   &03 06 17.8&17 18 20& 0.4169 &21.88 \\
A13   &03 06 16.7&17 18 25& 0.4201 &20.68 &B13   &03 06 08.6&17 19 44& 0.1624 &20.09 \\
A14   &03 06 18.3&17 18 31& 0.4209 &20.98 &B14   &03 06 12.3&17 18 25& 0.1842 &21.42 \\
A15   &03 06 20.3&17 18 22& 0.4176 &21.44 &B15   &03 06 14.8&17 17 20& 0.2453 &19.18 \\
A16   &03 06 23.4&17 18 35&	   &22.05 &B16   &03 06 10.7&17 17 55& 0.3078 &21.68 \\
A17   &03 06 24.8&17 18 05& 0.4069 &21.11 &B17   &03 06 11.9&17 17 07& 0.3095 &20.34 \\
A18   &03 06 24.8&17 17 48& 0.4192 &22.85 &B18   &03 06 05.5&17 17 57&     &20.91 \\
A19   &03 06 23.9&17 17 14&	   &21.69 &B19   &03 06 09.7&17 16 37& 0.5290 &22.45 \\
A4b   &03 06 12.8&17 22 35& 0.4190 &	  &	 &	    &	     &        &      \\
A4c   &03 06 12.2&17 22 26& 0.7463 &23.40 &	 &	    &	     &        &      \\
A14b  &03 06 19.0&17 18 20& 0.4171 &21.07 &	 &	    &	     &        &      \\
A17b  &03 06 24.0&17 18 18& 0.8060 &22.11 &	 &	    &	     &        &      \\
\hline
\end{tabular}
\end{table*}
\newpage
\begin{table*}
\centering
\caption{MOS targets in the field of \clvier\ observed with Setup A (PA$=-45^{\circ}$) and 
Setup B (PA$=+45^{\circ}$). A11$=$B9.}
\begin{tabular}{ccccc|ccccc}
\hline
ID    &RA        &DE         &z       &I	  &ID    &RA        &DE          &z     &I    \\
      &[h m s]   &[${\circ}$ ' '']&     &[mag] &      &[h m s]   &[${\circ}$ ' '']&     &[mag] \\	 
\hline
A1    &04 13 17.6&-65 48 46&	    &21.05  &B1    &04 12 24.9&-65 48 40&0.5080  &20.32  \\ 
A2    &04 13 11.8&-65 48 40&	    &20.60  &B2    &04 12 32.3&-65 48 24&	 &21.42   \\ 
A3    &04 13 06.6&-65 48 42&	    &20.53  &B3    &04 12 22.3&-65 49 53&	 &21.19   \\ 
A4    &04 13 02.1&-65 48 39&0.6537  &20.64  &B4    &04 12 29.7&-65 49 26&0.5525  &22.31   \\ 
A5    &04 13 02.8&-65 49 02&	    &22.19  &B5    &04 12 25.9&-65 50 40&0.6066  &20.20  \\ 
A6    &04 12 57.0&-65 49 06&	    &21.00  &B6    &04 12 34.3&-65 50 02&0.5082  &20.99  \\ 
A7    &04 12 54.5&-65 49 35&0.5644  &20.78  &B7    &04 12 37.8&-65 50 21&	 &19.87   \\ 
A8    &04 12 53.4&-65 49 48&0.4989  &21.59  &B8    &04 12 50.0&-65 49 49&0.6260  &21.05  \\ 
A9    &04 12 55.6&-65 50 33&0.8017  &20.94  &B9    &04 12 44.9&-65 50 38&0.5395  &19.68   \\ 
A10   &04 12 49.7&-65 50 45&	    &21.04  &B10   &04 12 49.4&-65 50 52&0.1847  &20.61   \\ 
A11   &04 12 44.9&-65 50 38&0.5395  &19.68  &B11   &04 12 52.4&-65 51 07&0.5099  &19.92  \\ 
A12   &04 12 40.8&-65 50 37&	    &21.67  &B12   &04 12 54.0&-65 51 23&0.5103  &21.15  \\ 
A13   &04 12 44.1&-65 51 37&0.6067  &21.01  &B13   &04 12 58.3&-65 51 16&0.5082  &20.39   \\ 
A14   &04 12 36.8&-65 51 23&0.5391  &19.86  &B14   &04 12 52.9&-65 52 27&0.5097  &21.47  \\ 
A15   &04 12 40.2&-65 52 14&0.5423  &20.07  &B15   &04 12 58.7&-65 52 22&0.6186  &21.22   \\ 
A16   &04 12 39.3&-65 52 40&	    &20.58  &B16   &04 13 03.3&-65 52 38&	 &	\\
A17   &04 12 36.3&-65 52 51&0.5381  &20.85  &B17   &04 13 08.0&-65 52 29&0.8479  &22.00 	\\ 
A18   &04 12 34.4&-65 53 17&0.6082  &20.87  &B18   &04 13 14.7&-65 52 23&	 &21.79 	\\
A19   &04 12 21.4&-65 52 15&	    &20.08  &B19   &04 13 12.5&-65 53 02&	 &21.40 	\\					     
A9b   &04 12 55.2&-65 50 35&0.4254  &19.18  &B8b   &04 12 49.7&-65 49 46&	 &21.71 	\\
A12b  &04 12 40.8&-65 50 36&0.0853  &16.70  &B10b  &04 12 49.4&-65 50 52&0.5035  &21.19    \\
A17b  &04 12 35.8&-65 52 54&0.5062  &21.38  &	   &	      & 	&\\
\hline
\end{tabular}
\end{table*}
\begin{table*}
\centering
\caption{All galaxies from which we obtained V$_{max}$. 
In the case of \mseins\ A12$=$B8.
For an identification see finding charts in fig.~3.
In some cases there was no [OII] line within the spectra and we used other emission lines.}
\begin{tabular}{ccccccccccccc}
\hline
cluster &ID     &mem    &Type  &incl. &$\delta$ &r$_{d}$  &V$_{max}$  &EW ([OII])      &M$_{B}$ \\
        &       &       &      &[deg] &[deg]    &['']     &[km/s]     &[\AA ]          &[mag] \\
(1)     & (2)   & (3)   &(4)   & (5)  & (6)     & (7)     & (8)       &(9)      & (10)  \\  
\hline
\mseins &A8     &yes    &+8 &45  &   5     &0.40     & 70$\pm$30 &72.8   &-20.19  \\
\mseins &A12    &yes    &+5 &50  &  43     &0.70     &220$\pm$59 &	  &-20.69   \\
\mseins &A15    &yes    &+5 &20  &  20     &0.57     &270$\pm$126&	  &-21.24   \\
\mseins &B1     &yes    &+5 &60  &  -9     &1.35     &130$\pm$23 &	  &-20.15   \\
\mseins &B3     &no     &+1 &55  &  63     &0.40     &320$\pm$129&5.1    &-21.26   \\
\mseins &B5     &yes    &+5 &60  &  10     &0.60     &150$\pm$26 &	  &-20.32   \\
\mseins &B8     &yes    &+5 &50  & -17     &0.70     &240$\pm$49 &	  &-20.69   \\
\mseins &B12    &yes    &+5 &50  &   8     &0.60     &160$\pm$32 &	  &-20.52   \\
\mseins &B14    &yes    &+5 &52  & -50     &1.02     &230$\pm$68 &15.7    &-21.48   \\
\hline 
\cldrei &B4     &yes    &+3 &20  & -30     &0.80     &400$\pm$135&5.1    &-22.13     \\
\cldrei &B7     &no     &+5 &35  &  25     &0.72     &320$\pm$62 &19.4   &-21.98     \\
\hline 
\clvier &A7     &no  	&+3 &54  & 19	   &0.45     &200$\pm$26 &12.3   &-20.94  \\
\clvier &A8     &yes 	&+8 &65  &-20	   &0.60     &150$\pm$18 &39.5   &-20.12  \\
\clvier &A11    &no  	&+3 &33  &  0	   &0.85     &350$\pm$96 &10.3   &-21.65  \\
\clvier &A13    &no  	&+5 &50  & 25	   &0.55     &202$\pm$29 &14.1   &-21.16  \\
\clvier &A18    &no  	&+5 &60  & 20	   &0.80     &205$\pm$38 &36.1   &-21.30  \\
\clvier &B5     &no  	&+5 &35  & -5	   &0.95     &300$\pm$79 &25.3   &-21.53  \\
\clvier &B6     &yes 	&+5 &40  &-15	   &0.25     &165$\pm$27 &17.1   &-20.41  \\
\clvier &B11    &yes 	&+3 &55  & 15	   &0.52     &180$\pm$23 &11.4   &-21.52  \\
\clvier &B12    &yes 	&+5 &43  &-75	   &0.30     &230$\pm$130&7.2	 &-20.42  \\
\clvier &B14    &yes 	&+3 &63  &-22	   &0.40     &220$\pm$27 &4.5	 &-20.00  \\
\hline 
\end{tabular}
\end{table*}
\begin{figure*}
\centerline{\hbox{
\psfig{figure=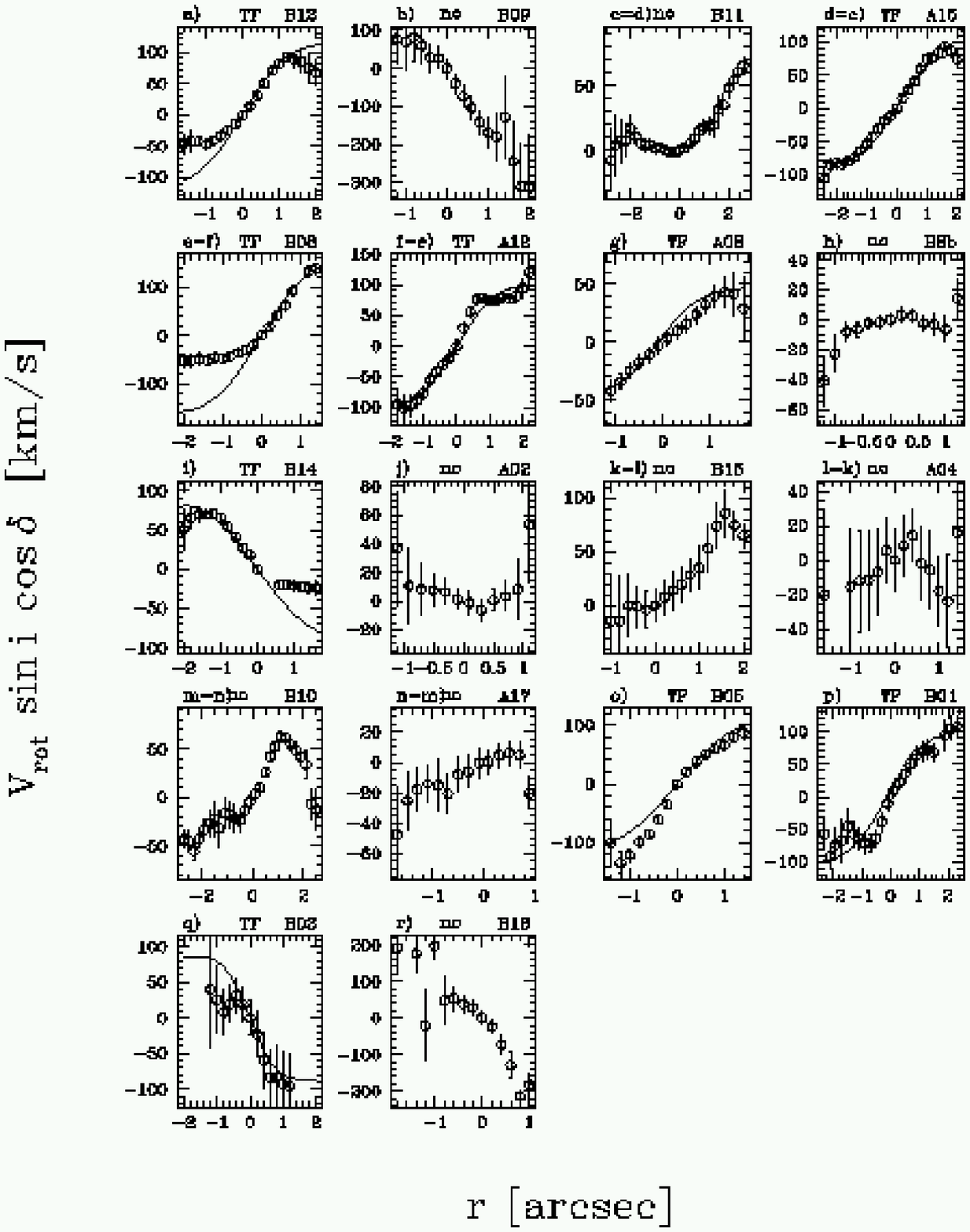,clip=t}
}}
\caption {Position velocity diagrams of galaxies in the field of \mseins .
Galaxies which entered the TF diagram presented in \science ~are labeled
(TF). The solid line represents the projected best fitting model profile. See Table 6--8 for
comparison.}
\end{figure*}
\begin{figure*}
\centerline{\hbox{
\psfig{figure=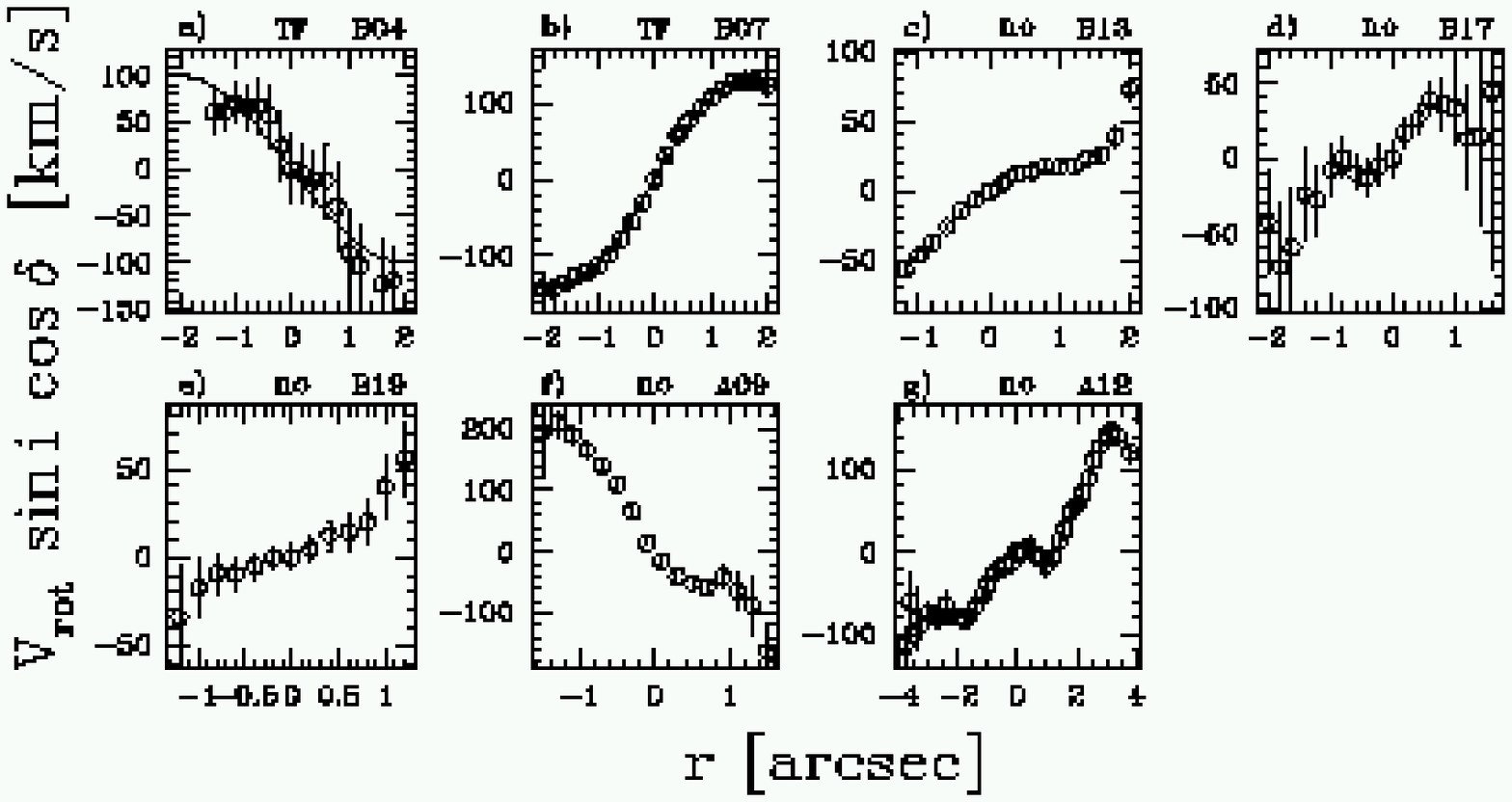,clip=t}
}}
\caption {Position velocity diagrams of galaxies in the field of \cldrei .
Galaxies which entered the TF diagram presented in \science ~are labeled
(TF). The solid line represents the projected best fitting model profile.
See Table 9 for comparison.}
\end{figure*}
\begin{figure*}
\centerline{\hbox{
\psfig{figure=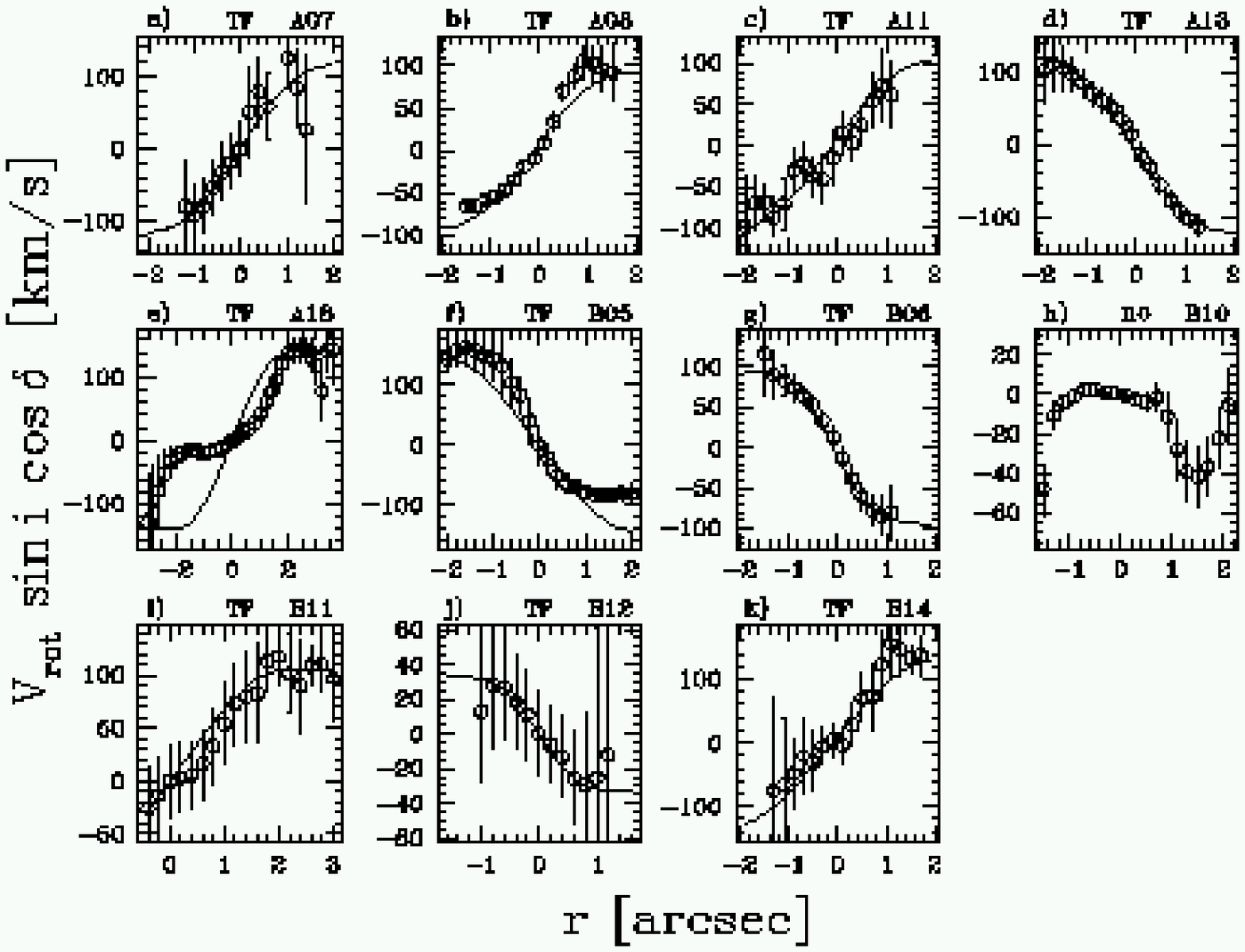,clip=t}
}}
\caption {Position velocity diagrams of galaxies in the field of \clvier .
Galaxies which entered the TF diagram presented in \science ~are labeled
(TF). The solid line represents the projected best fitting model profile.
See Table 10 and 11 for comparison.}
\end{figure*}
\begin{table*}[h]
\caption{Measured rotation velocities V$_{rot}$ as function of projected distance (Pos) to 
the galaxies' centers (Pos$=$0) for \mseins . Galaxies are indicated by their ID with the measured emission line, respectively. 
See Fig.4 for comparison.}
\begin{center}
\begin{tabular}{ccccccccc}
\hline     
\multicolumn{1}{c}{Pos} &\multicolumn{8}{c}{V$_{rot}$ sin $i$ cos $\delta $}\\
\multicolumn{1}{c}{['']}&\multicolumn{8}{c}{[km s$^{-1}$]}\\
\hline
&B12(H$_{\beta}$)&B9(O[II])&B11(H$_{\beta}$)&A15(O[III])&B8(H$_{\beta}$)&A12(O[III])&B8b(O[III])&B14(H$_{\beta}$)\\
\hline
-2.8&	           &	            &  -7.1$\pm$24.0&		     &  	      & 	       &	       &      \\
-2.6&	           &	            &  +4.3$\pm$23.3&		     &  	      & 	       &	       &      \\
-2.4&	           &	            &  +7.7$\pm$16.8&-107.2$\pm$19.1&		     &  	      & 	      &      \\
-2.2&	           &	            &  +7.8$\pm$12.3& -85.1$\pm$9.8 &		   &		    &		    &	   \\
-2.0&	           &	            & +17.4$\pm$9.5 & -82.6$\pm$8.3 & -51.3$\pm$13.8&		   &		  & +50.3$\pm$25.1\\
-1.8&	           &	            & +11.1$\pm$5.9 & -85.2$\pm$7.3 & -51.3$\pm$14.3& -95.7$\pm$26.5&		  & +65.2$\pm$13.2\\
-1.6&-47.3$\pm$16.0&	            &  +6.0$\pm$5.2 & -79.7$\pm$5.5 & -48.3$\pm$13.7&-100.1$\pm$17.9&		  & +69.5$\pm$8.6\\
-1.4&-42.7$\pm$16.4&	            &  +4.3$\pm$4.5 & -74.4$\pm$5.6 & -51.3$\pm$11.2& -97.8$\pm$16.4&		  & +69.6$\pm$5.6\\
-1.2&-42.7$\pm$7.9 & +76.6$\pm$64.4 &  +4.2$\pm$4.2 & -64.1$\pm$6.1 & -45.4$\pm$10.7& -88.0$\pm$12.7&-40.8$\pm$16.6& +71.1$\pm$3.3\\
-1.0&-47.6$\pm$7.1 & +69.4$\pm$75.5 &  +2.1$\pm$3.8 & -54.9$\pm$6.1 & -46.7$\pm$9.1 & -76.2$\pm$10.7&-22.8$\pm$12.0& +65.6$\pm$4.4\\
-0.8&-40.9$\pm$6.2 & +80.8$\pm$61.0 &  +2.1$\pm$3.5 & -42.8$\pm$6.4 & -43.2$\pm$8.1 & -55.8$\pm$10.1& -8.1$\pm$6.8 & +53.3$\pm$5.1\\
-0.6&-33.6$\pm$5.3 & +59.5$\pm$48.4 &  +0.8$\pm$2.9 & -30.4$\pm$5.9 & -36.4$\pm$8.0 & -42.2$\pm$8.4 & -6.4$\pm$5.4 & +41.5$\pm$5.5\\
-0.4&-24.7$\pm$4.5 & +28.5$\pm$35.9 &  -0.4$\pm$2.5 & -18.7$\pm$5.8 & -29.1$\pm$7.8 & -30.1$\pm$9.6 & -2.3$\pm$4.0 & +27.4$\pm$5.9\\
-0.2&-13.8$\pm$4.3 & +27.6$\pm$27.9 &  -0.4$\pm$2.3 & -12.2$\pm$4.7 & -18.8$\pm$7.5 & -18.8$\pm$10.4& -1.6$\pm$3.8 & +17.6$\pm$5.5\\
+0.0& +0.0$\pm$4.2 &  +0.0$\pm$26.6 &  +0.0$\pm$2.2 &  +0.0$\pm$4.7 &  +0.0$\pm$8.0 &  +0.0$\pm$12.2& +0.0$\pm$4.4 &  +0.0$\pm$5.2\\
+0.2&+14.1$\pm$4.1 & -39.2$\pm$25.0 &  +1.9$\pm$2.4 & +14.9$\pm$6.6 & +18.3$\pm$8.9 & +29.0$\pm$11.1& +3.3$\pm$4.8 &    \\
+0.4&+29.9$\pm$4.0 & -73.9$\pm$28.0 &  +5.5$\pm$2.8 & +27.8$\pm$6.8 & +41.5$\pm$9.6 & +55.8$\pm$9.2 & +2.3$\pm$4.4 &    \\
+0.6&+49.9$\pm$5.0 &-103.3$\pm$30.2 &  +9.2$\pm$4.1 & +40.5$\pm$6.9 & +62.6$\pm$11.2& +75.7$\pm$6.7 & -2.4$\pm$5.3 & -20.7$\pm$3.4\\
+0.8&+69.7$\pm$6.2 &-141.0$\pm$30.7 & +15.7$\pm$5.0 & +59.7$\pm$8.4 & +92.3$\pm$10.4& +78.7$\pm$5.2 & -3.2$\pm$6.8 & -19.0$\pm$3.8\\
+1.0&+81.9$\pm$5.5 &-168.5$\pm$41.4 & +18.2$\pm$5.9 & +75.2$\pm$8.5 &+132.3$\pm$12.9& +74.1$\pm$4.8 & -6.5$\pm$8.6 & -21.2$\pm$3.8\\
+1.2&+92.5$\pm$6.0 &-179.7$\pm$64.9 & +18.9$\pm$6.8 & +77.4$\pm$8.0 &+138.3$\pm$11.7& +74.9$\pm$5.0 &+14.3$\pm$13.0& -21.7$\pm$4.5\\
+1.4&+87.4$\pm$11.4&-127.2$\pm$108.3& +19.8$\pm$8.8 & +84.9$\pm$7.2 &		   & +79.4$\pm$6.1 &		  & -23.8$\pm$5.9\\
+1.6&+85.8$\pm$19.5&-244.6$\pm$102.0& +32.1$\pm$9.7 & +93.8$\pm$6.8 &		   & +76.6$\pm$6.9 &		  & -23.9$\pm$10.3\\
+1.8&+73.7$\pm$24.3&-312.3$\pm$112.9& +35.3$\pm$8.2 & +87.3$\pm$9.1 &		   & +81.2$\pm$10.7&		  &	 \\
+2.0&+67.3$\pm$21.5&-313.4$\pm$137.8& +48.3$\pm$6.8 & +74.5$\pm$12.1&		    & +93.7$\pm$16.3&		    &	   \\
+2.2&	           &	            & +54.6$\pm$5.1 &		    &		    &+120.6$\pm$29.0&	     &      \\
+2.4&	           &	            & +61.3$\pm$5.4 &		    &		    &  	      & 	      &      \\
+2.6&	           &	            & +65.9$\pm$6.6 &		    &		    &  	      & 	      &      \\
\hline  															    
\end{tabular}
\end{center}
\end{table*}
---------------------------------------------------------------------------------------------------------------------------------------------
\begin{table*}
\caption{Measured rotation velocities V$_{rot}$ as function of projected distance (Pos) to 
the galaxies' centers (Pos$=$0) for \mseins . Galaxies are indicated by their ID with the measured emission line, respectively. 
See Fig.4 for comparison.}
\begin{center}
\begin{tabular}{ccccccc}
\hline
\multicolumn{1}{c}{Pos} &\multicolumn{5}{c}{V$_{rot}$ sin $i$ cos $\delta $}\\
\multicolumn{1}{c}{['']}&\multicolumn{5}{c}{[km s$^{-1}$]}\\
\hline
&B15(O[II])&A4(O[II])&B5(H$_{\beta}$)&B3(O[II])&B16(O[II])\\
\hline     
-1.8&                &               &                &                &+191.4$\pm$72.2\\
-1.6&                &-20.4$\pm$49.4 &                &                &           	\\
-1.4&                &               & -99.6$\pm$40.2&  	      &+175.7$\pm$51.8\\
-1.2&                &               &-134.5$\pm$19.1& +39.9$\pm$80.4& -21.0$\pm$98.0\\
-1.0& -14.3$\pm$41.3 &-15.5$\pm$33.8 &-121.1$\pm$10.8& +25.9$\pm$47.2&+197.0$\pm$34.2\\
-0.8& -14.1$\pm$41.9 &-12.2$\pm$29.4 & -98.8$\pm$8.3 &  +8.1$\pm$32.4& +47.5$\pm$66.0\\
-0.6&  +0.1$\pm$28.7 &-11.6$\pm$28.4 & -84.7$\pm$6.3 & +19.6$\pm$26.3& +52.7$\pm$33.4\\
-0.4&  -1.4$\pm$19.6 & -6.9$\pm$25.3 & -60.6$\pm$6.5 & +30.9$\pm$24.6& +37.6$\pm$24.7\\
-0.2&  -4.4$\pm$16.9 & +5.6$\pm$18.5 & -34.7$\pm$7.7 & +17.8$\pm$24.5& +28.5$\pm$21.5\\
+0.0&  +0.0$\pm$15.4 & +0.0$\pm$18.3 &  +0.0$\pm$6.3 &  +0.0$\pm$24.0&  +0.0$\pm$18.8\\
+0.2&  +7.6$\pm$16.4 & +8.3$\pm$17.2 & +21.4$\pm$6.4 & -23.0$\pm$31.6& -24.6$\pm$17.4\\
+0.4& +14.0$\pm$17.8 &+14.3$\pm$14.7 & +37.3$\pm$5.1 & -58.8$\pm$38.8& -72.4$\pm$27.3\\
+0.6& +19.7$\pm$15.6 & -1.9$\pm$21.6 & +49.3$\pm$5.1 & -83.6$\pm$40.0&-129.0$\pm$37.2\\
+0.8& +28.2$\pm$14.7 & -5.8$\pm$22.3 & +59.1$\pm$4.3 & -83.4$\pm$43.4&-214.7$\pm$16.1\\
+1.0& +35.2$\pm$20.4 &-18.0$\pm$20.6 & +65.8$\pm$6.1 & -93.0$\pm$46.0&-183.8$\pm$29.6\\
+1.2& +53.2$\pm$22.4 &-24.0$\pm$26.8 & +79.5$\pm$6.5 & -95.4$\pm$44.7&  	   \\
+1.4& +73.6$\pm$21.3 &+16.1$\pm$43.9 & +87.0$\pm$10.2&  	    &		     \\
+1.6& +85.8$\pm$22.4&		    &		     &  	      & 	       \\
+1.8& +74.4$\pm$12.8&		    &		     &  	      & 	       \\
+2.0& +64.9$\pm$11.0&		    &		     &  	      & 	       \\
\hline  															    
\end{tabular}
\end{center}
\end{table*}
\begin{table*}
\caption{Measured rotation velocities V$_{rot}$ as function of projected distance (Pos) to 
the galaxies' centers (Pos$=$0) for \mseins . Galaxies are indicated by their ID with the measured emission line, respectively. 
See Fig.4 for comparison.}
\begin{center}
\begin{tabular}{cccccc}
\hline
\multicolumn{1}{c}{Pos} &\multicolumn{5}{c}{V$_{rot}$ sin $i$ cos $\delta $}\\
\multicolumn{1}{c}{['']}&\multicolumn{5}{c}{[km s$^{-1}$]}\\
\hline
&A8(O[II])&A2(O[III])&B10(H$_{\beta}$)&A17(O[III])&B1(H$_{\beta}$)\\
\hline     
-2.7&	              &                & -43.2$\pm$+9.3&		 &		  \\
-2.5&	              &                & -45.1$\pm$11.7&		 &		  \\
-2.3&	              &                & -55.8$\pm$8.7 &		 & -57.7$\pm$24.2\\
-2.1&	              &                & -43.3$\pm$7.1 &		 & -91.7$\pm$19.4\\
-1.9&	              &                & -33.3$\pm$10.0&		 & -74.0$\pm$24.9\\
-1.7&	              &                & -26.2$\pm$13.1& -47.6$\pm$31.0& -67.5$\pm$33.1\\
-1.5&	              &                & -22.5$\pm$17.9& -25.1$\pm$18.9& -45.4$\pm$27.3\\
-1.3&	              &                & -32.0$\pm$19.1& -17.6$\pm$14.1& -57.4$\pm$20.5\\
-1.1& -42.1$\pm$8.4 & +36.6$\pm$40.1& -16.6$\pm$15.5& -13.9$\pm$11.7& -63.4$\pm$16.7\\
-0.9& -34.0$\pm$6.7 & +10.7$\pm$25.7& -17.4$\pm$13.3& -14.9$\pm$17.1& -72.0$\pm$15.9\\
-0.7& -24.6$\pm$6.6 &  +8.1$\pm$18.3& -21.5$\pm$14.7& -20.4$\pm$13.3& -71.4$\pm$14.2\\
-0.5& -17.4$\pm$6.2 &  +7.5$\pm$11.3& -23.8$\pm$13.1&  -7.8$\pm$11.2& -64.3$\pm$13.7\\
-0.3& -10.8$\pm$6.6 &  +6.2$\pm$9.0 & -12.8$\pm$10.7&  -5.6$\pm$  9.9& -39.1$\pm$14.5\\
-0.1&  -3.2$\pm$6.7 &  +1.4$\pm$7.1 &  -4.0$\pm$6.3&  -0.2$\pm$  9.4& -10.2$\pm$12.9\\
+0.1&  +3.2$\pm$6.7 &  -1.4$\pm$8.1 &  +4.0$\pm$5.3&  +0.2$\pm$  7.7& +10.2$\pm$11.4\\
+0.3&  +9.4$\pm$6.7 &  -6.4$\pm$7.2 & +10.4$\pm$4.7&  +4.5$\pm$  7.7& +22.8$\pm$12.0\\
+0.5& +15.6$\pm$7.3 &  +0.5$\pm$7.3 & +26.5$\pm$4.6&  +6.1$\pm$  8.0& +35.1$\pm$13.2\\
+0.7& +23.7$\pm$7.4 &  +3.3$\pm$10.6& +42.1$\pm$5.3&  +4.9$\pm$  8.6& +51.2$\pm$13.3\\
+0.9& +32.2$\pm$8.1 &  +8.3$\pm$21.4& +52.0$\pm$5.4& -19.8$\pm$ 10.9& +63.6$\pm$14.5\\
+1.1& +39.5$\pm$9.6 & +54.0$\pm$41.2& +62.7$\pm$5.2&		    & +69.1$\pm$14.6\\
+1.3& +43.0$\pm$13.0&		     & +60.9$\pm$6.3&		      & +71.2$\pm$14.8\\
+1.5& +41.3$\pm$18.3&		     & +54.8$\pm$5.6&		      & +67.9$\pm$12.9\\
+1.7& +28.9$\pm$27.2&		     & +49.1$\pm$7.5&		      & 	       \\
+1.9&	              &                & +42.3$\pm$8.2 &		 & +94.2$\pm$22.0\\
+2.1&	              &                & +34.2$\pm$14.1&		 &+105.1$\pm$19.2\\
+2.3&	              &                &  -6.6$\pm$15.3&		 &+106.8$\pm$16.0\\
+2.5&	              &                & -12.5$\pm$21.2&		 &		  \\
\hline  															    
\end{tabular}
\end{center}
\end{table*}
----------------------------------------------------------------------
\begin{table*}
\caption{Measured rotation velocities V$_{rot}$ as function of projected distance (Pos) to 
the galaxies' centers (Pos$=$0) for \cldrei . Galaxies are indicated by their ID with the measured emission line, respectively. 
See Fig.5 for comparison.}
\begin{center}
\begin{tabular}{ccccccc|cc}
\hline
\multicolumn{1}{c}{Pos} &\multicolumn{6}{c}{V$_{rot}$ sin $i$ cos $\delta $} &\multicolumn{1}{c}{Pos} &\multicolumn{1}{c}{V$_{rot}$ sin $i$ cos $\delta $}\\
\multicolumn{1}{c}{['']}&\multicolumn{6}{c}{[km s$^{-1}$]} &\multicolumn{1}{c}{['']}&\multicolumn{1}{c}{[km s$^{-1}$]}\\
\hline
&B4(O[III])&B7(O[II])&B13(O[III])&B17(H$_{\beta}$)&B19(O[II])&A12(O[III])&&A9(O[II])\\
\hline     
-3.8&                &                &               &                &           	&-112.9$\pm$21.8&&\\
-3.6&                &                &               &                &           	& -60.1$\pm$36.2&&\\
-3.4&                &                &               &                &           	& -80.9$\pm$37.3&&\\
-3.2&                &                &               &                &           	& -87.4$\pm$18.2&&\\
-3.0&                &                &               &                &           	& -74.0$\pm$14.1&&\\
-2.8&                &                &               &                &           	& -78.7$\pm$14.6&&\\
-2.6&                &                &               &                &           	& -79.0$\pm$14.5&&\\
-2.4&                &                &               &                &           	& -63.5$\pm$16.0&&\\
-2.2&                &                &               &                &           	& -78.2$\pm$8.9&&\\
-2.0&                &-142.9$\pm$19.3&  	     & -42.2$\pm$33.1&  	      & -79.0$\pm$8.5&&\\
-1.8&                &-143.2$\pm$13.4&  	     & -72.7$\pm$42.3&  	      & -82.9$\pm$7.6&&\\
-1.6&                &-136.1$\pm$9.3&		    & -59.7$\pm$39.8&		     & -76.8$\pm$8.1&-1.5&+195.$\pm$71.3\\
-1.4& +60.7$\pm$23.7&-126.4$\pm$7.8&		   & -23.9$\pm$31.2&		    & -66.4$\pm$9.3&-1.3&+208.$\pm$22.4\\
-1.2& +60.7$\pm$20.5&-120.5$\pm$6.8& -56.6$\pm$3.3& -28.1$\pm$23.7& -33.9$\pm$28.5& -52.7$\pm$10.3&-1.1&+190.$\pm$12.9\\
-1.0& +69.6$\pm$21.4&-113.9$\pm$6.3& -46.3$\pm$2.8&  -8.0$\pm$16.9& -17.3$\pm$16.4& -39.6$\pm$11.6&-0.9&+165.$\pm$10.8\\
-0.8& +65.8$\pm$23.8& -98.7$\pm$6.0& -37.0$\pm$2.3&  -0.2$\pm$13.3&  -9.0$\pm$11.5& -25.3$\pm$11.6&-0.7&+140.$\pm$11.4\\
-0.6& +67.1$\pm$29.6& -78.2$\pm$5.0& -25.5$\pm$1.7&  -9.7$\pm$10.4&  -9.7$\pm$9.7& -20.9$\pm$11.9&-0.5&+109.$\pm$9.3\\
-0.4& +56.5$\pm$34.1& -55.1$\pm$4.3& -14.1$\pm$1.4& -13.4$\pm$12.1&  -5.4$\pm$7.9& -16.6$\pm$12.2&-0.3& +65.$\pm$5.4\\
-0.2& +24.4$\pm$35.4& -29.3$\pm$5.0&  -6.1$\pm$1.9&  -4.8$\pm$13.0&  -0.3$\pm$6.9& -12.5$\pm$13.2&-0.1& +15.$\pm$4.8\\
+0.0&  +0.0$\pm$36.3&  +0.0$\pm$5.8&  +0.0$\pm$2.5&  +0.0$\pm$11.8&  +0.0$\pm$7.0&  +0.0$\pm$11.8&+0.1& -15.$\pm$5.0\\
+0.2&  -6.3$\pm$30.8& +30.8$\pm$5.8&  +5.9$\pm$3.0& +17.3$\pm$9.6&  +4.5$\pm$7.1&  +1.1$\pm$11.6&+0.3& -40.$\pm$4.9\\
+0.4& -11.2$\pm$28.2& +57.6$\pm$5.5& +11.5$\pm$2.8& +26.1$\pm$10.1& +12.0$\pm$8.4&  +5.3$\pm$12.6&+0.5& -51.$\pm$5.7\\
+0.6& -12.2$\pm$39.9& +76.5$\pm$6.2& +13.9$\pm$3.1& +39.1$\pm$11.1& +14.6$\pm$10.4&  -5.2$\pm$10.7&+0.7& -59.$\pm$7.5\\
+0.8& -38.9$\pm$46.9& +93.3$\pm$8.1& +18.0$\pm$3.4& +36.1$\pm$16.0& +19.7$\pm$12.7&  -7.4$\pm$12.0&+1.9& -41.$\pm$17.8\\
+1.0& -89.3$\pm$45.7&+106.0$\pm$9.0& +17.5$\pm$3.6& +33.5$\pm$25.3& +39.9$\pm$17.0& -15.5$\pm$12.4&+1.1& -64.$\pm$31.8\\
+1.2&-103.9$\pm$43.2&+117.7$\pm$8.6& +18.6$\pm$3.6& +13.1$\pm$34.1& +56.0$\pm$20.7&  -6.1$\pm$13.1&+1.3& -87.$\pm$46.2\\
+1.4&               &+127.3$\pm$7.1& +24.1$\pm$4.3& +14.7$\pm$59.4&		   & +13.3$\pm$11.9&+1.5&-163.$\pm$44.5\\
+1.6&-122.5$\pm$48.8&+129.2$\pm$8.5& +26.3$\pm$5.7& +43.8$\pm$118.0&		    & +27.5$\pm$12.5&&\\
+1.8&-119.3$\pm$37.5&+126.0$\pm$10.8& +39.6$\pm$7.9&		    &		     & +48.8$\pm$12.4&&\\
+2.0&	             &+122.4$\pm$14.0& +72.9$\pm$7.3&		     &  	      & +56.7$\pm$13.0&&\\
+2.2&	             &                &               &                &           	& +72.0$\pm$13.1&&\\
+2.4&	             &                &               &                &           	& +89.3$\pm$13.7&&\\
+2.6&	             &                &               &                &           	&+112.1$\pm$13.6&&\\
+2.8&	             &                &               &                &           	&+128.6$\pm$11.4&&\\
+3.0&	             &                &               &                &           	&+139.5$\pm$9.2&&\\
+3.2&	             &                &               &                &           	&+147.8$\pm$7.6&&\\
+3.4&	             &                &               &                &           	&+141.2$\pm$8.5&&\\
+3.6&	             &                &               &                &           	&               &&\\
+3.8&	             &                &               &                &           	&+120.5$\pm$11.0&&\\
\hline																  
\end{tabular}
\end{center}
\end{table*}

\begin{table*}
\caption{Measured rotation velocities V$_{rot}$ as function of projected distance (Pos) to 
the galaxies' centers (Pos$=$0) for \clvier . Galaxies are indicated by their ID with the measured emission line, respectively. 
See Fig.6 for comparison.}
\begin{center}
\begin{tabular}{cccccc}
\hline
\multicolumn{1}{c}{Pos} &\multicolumn{4}{c}{V$_{rot}$ sin $i$ cos $\delta $}\\
\multicolumn{1}{c}{['']}&\multicolumn{4}{c}{[km s$^{-1}$]}\\
\hline
&A7(O[II])&A18(O[II])&B5(O[II])&B11(O[II])&B12(O[II])\\
\hline     
-3.0&&-130.3$\pm$80.2&  	       &		&		  \\
-2.8&&-115.7$\pm$75.9&  	       &		&		  \\
-2.6&& -78.4$\pm$55.3&  	       &		&		  \\
-2.4&& -51.2$\pm$35.9&  	       &		&		  \\
-2.2&& -31.6$\pm$23.0&  	       &		&		  \\
-2.0&& -26.3$\pm$15.1&+142.9$\pm$19.7&  	      & 		\\
-1.8&& -17.6$\pm$10.5&+156.5$\pm$21.8&  	      & 		\\
-1.6&& -14.8$\pm$8.5&+161.6$\pm$26.0&  	      & 		\\
-1.4&& -15.2$\pm$8.4&+159.4$\pm$31.6&  	      & 		\\
-1.2&-79.2$\pm$64.7& -16.3$\pm$8.8&+150.4$\pm$33.3&		   &		     \\
-1.0&-93.0$\pm$44.9& -18.1$\pm$8.8&+145.4$\pm$39.3&		   & +12.7$\pm$39.7\\
-0.8&-79.5$\pm$38.3& -16.3$\pm$9.5&+130.6$\pm$46.7&		   & +28.1$\pm$36.5\\
-0.6&-53.7$\pm$38.7& -13.1$\pm$10.5&+104.4$\pm$38.3&		   & +26.7$\pm$30.1\\
-0.4&-32.6$\pm$39.8& -11.7$\pm$11.0& +82.8$\pm$31.4& -25.9$\pm$38.3& +18.4$\pm$26.1\\
-0.2&-18.1$\pm$37.5&  -6.9$\pm$11.3& +39.1$\pm$35.0& -13.1$\pm$36.7& +11.2$\pm$26.3\\
+0.0& +0.0$\pm$39.1&  +0.0$\pm$11.5&  +0.0$\pm$32.4&  +0.0$\pm$34.2&  +0.0$\pm$23.8\\
+0.2&+50.9$\pm$62.4&  +7.2$\pm$11.1& -22.4$\pm$26.5&  +3.6$\pm$33.7&  -6.1$\pm$21.9\\
+0.4&+78.9$\pm$48.0& +14.9$\pm$11.8& -50.9$\pm$22.0&  +5.6$\pm$31.4& -13.0$\pm$22.9\\
+0.6&+58.7$\pm$45.3& +19.9$\pm$12.9& -64.9$\pm$15.2& +16.3$\pm$32.7& -25.2$\pm$26.7\\
+0.8&		  & +27.6$\pm$14.7& -70.8$\pm$12.2& +33.2$\pm$36.0& -29.5$\pm$41.9\\
+1.0&+125.3$\pm$39.6& +42.3$\pm$16.5& -78.0$\pm$9.6& +54.6$\pm$39.1& -25.3$\pm$86.9\\
+1.2& +84.5$\pm$53.7& +60.5$\pm$18.1& -82.8$\pm$9.4& +73.3$\pm$38.5& -12.5$\pm$166.9\\
+1.4& +26.9$\pm$102.3& +78.1$\pm$20.1& -83.6$\pm$10.3& +79.7$\pm$42.8&  	      \\
+1.6&& +97.2$\pm$21.7& -84.8$\pm$11.2& +83.1$\pm$46.9&  	       \\
+1.8&&+122.5$\pm$21.5& -82.0$\pm$12.7&+112.9$\pm$27.9&  	       \\
+2.0&&+139.5$\pm$19.8& -81.6$\pm$15.5&+117.5$\pm$26.3&  	       \\
+2.2&&+142.4$\pm$17.9&  	       &+101.7$\pm$38.0&		 \\
+2.4&&+142.5$\pm$17.9&  	       & +90.2$\pm$44.0&		 \\
+2.6&&+145.2$\pm$19.0&  	       &+109.1$\pm$19.1&		 \\
+2.8&&+141.2$\pm$20.6&  	       &+110.4$\pm$24.6&		 \\
+3.0&&+120.5$\pm$28.0&  	       & +97.5$\pm$40.2&		 \\
+3.2&& +78.4$\pm$44.1&  	       &		&		  \\
+3.4&&+141.8$\pm$33.5&  	       &		&		  \\
+3.6&&+143.5$\pm$54.8&  	       &		&		  \\
\hline							        								       
\end{tabular}
\end{center}
\end{table*}
\begin{table*}
\caption{Measured rotation velocities V$_{rot}$ as function of projected distance (Pos) to 
the galaxies' centers (Pos$=$0) for \clvier . Galaxies are indicated by their ID with the measured emission line, respectively. 
See Fig.6 for comparison.}
\begin{center}
\begin{tabular}{ccccccc}
\hline
\multicolumn{1}{c}{Pos} &\multicolumn{6}{c}{V$_{rot}$ sin $i$ cos $\delta $}\\
\multicolumn{1}{c}{['']}&\multicolumn{6}{c}{[km s$^{-1}$]}\\
\hline
&A8(O[II])&A11(O[III])&A13(O[II])&B6(O[II])&B10(O[III])&B14(O[II])\\
\hline     
-1.9&                 & -98.2$\pm$60.3&+102.2$\pm$46.5& 	       &		 &		   \\
-1.7&                 & -69.9$\pm$33.1&+112.5$\pm$39.3& 	       &		 &		   \\
-1.5& -65.4$\pm$7.2& -68.4$\pm$22.5&+104.7$\pm$30.9&+114.7$\pm$51.1& -48.1$\pm$12.9&		     \\
-1.3& -65.5$\pm$4.1& -87.1$\pm$24.3& +96.3$\pm$22.3& +89.1$\pm$29.2& -11.0$\pm$6.3& -75.5$\pm$146.5\\
-1.1& -57.7$\pm$3.9& -70.9$\pm$30.0& +84.5$\pm$18.3& +82.4$\pm$19.6&  -4.4$\pm$4.0& -66.3$\pm$103.3\\
-0.9& -53.4$\pm$3.9& -30.7$\pm$28.3& +70.8$\pm$16.0& +73.3$\pm$15.3&  -1.0$\pm$2.7& -48.8$\pm$56.0\\
-0.7& -46.4$\pm$4.0& -22.2$\pm$24.4& +61.1$\pm$17.2& +62.8$\pm$12.1&  +1.9$\pm$2.0& -20.8$\pm$60.2\\
-0.5& -34.5$\pm$4.2& -37.9$\pm$24.4& +50.7$\pm$18.3& +52.4$\pm$10.4&  +2.3$\pm$1.8& -30.2$\pm$47.2\\
-0.3& -17.6$\pm$4.3& -40.6$\pm$28.3& +34.9$\pm$17.6& +33.2$\pm$11.0&  +0.8$\pm$1.6&  -6.2$\pm$28.8\\
-0.1&  -8.6$\pm$4.8& -15.0$\pm$32.2& +13.9$\pm$17.4& +12.8$\pm$13.1&  +0.3$\pm$1.7&  +3.5$\pm$26.0\\
+0.1&  +8.6$\pm$5.4& +15.0$\pm$26.3& -13.9$\pm$15.9& -12.8$\pm$13.3&  -0.3$\pm$1.9&  -3.5$\pm$29.6\\
+0.3& +34.3$\pm$9.4&  +3.8$\pm$23.5& -32.1$\pm$13.3& -38.8$\pm$15.5&  -3.0$\pm$2.7& +28.0$\pm$33.8\\
+0.5& +70.3$\pm$10.2& +25.2$\pm$23.5& -55.3$\pm$11.8& -60.6$\pm$17.1&  -4.0$\pm$3.9& +69.8$\pm$37.8\\
+0.7& +86.6$\pm$14.9& +53.9$\pm$33.3& -76.8$\pm$11.7& -76.9$\pm$19.7&  -1.7$\pm$7.0& +73.2$\pm$52.3\\
+0.9&+104.3$\pm$18.6& +72.6$\pm$44.1& -91.5$\pm$12.4& -84.9$\pm$24.1& -11.2$\pm$12.1&+122.6$\pm$54.3\\
+1.1&+101.9$\pm$22.2& +62.4$\pm$40.3&-103.2$\pm$11.2& -80.1$\pm$33.9& -27.6$\pm$15.1&+155.1$\pm$51.6\\
+1.3& +96.1$\pm$26.9&		     &-111.8$\pm$13.4&  	      & -39.1$\pm$15.5&+146.1$\pm$39.9\\
+1.5& +92.3$\pm$32.7&		     &  	       &		& -41.7$\pm$15.5&+131.9$\pm$21.1\\
+1.7&                 &                &           	 &           	  & -36.4$\pm$14.0&+136.7$\pm$30.9\\
+1.9&                 &                &           	 &           	  & -22.0$\pm$15.7&		   \\
+2.1&                 &                &           	 &           	  &  -5.6$\pm$18.9&		    \\
\hline  															    
\end{tabular}
\end{center}
\end{table*}

\end{document}